\documentclass[12pt]{spieman}  % 12pt font required by SPIE;
\usepackage{amsmath,amsfonts,amssymb}
\usepackage{graphicx}
\usepackage{setspace}
\usepackage{tocloft}
\usepackage{caption}
\usepackage{subcaption}
\usepackage{tablefootnote}
\usepackage{verbatim}
\usepackage{multirow}
\usepackage{wrapfig}
\usepackage{tabularx}
\newcolumntype{L}[1]{>{\raggedright\arraybackslash}p{#1}}
\newcolumntype{C}[1]{>{\centering\arraybackslash}p{#1}}
\newcolumntype{R}[1]{>{\raggedleft\arraybackslash}p{#1}}

\usepackage[export]{adjustbox}

\usepackage{soul}

\title{The Keck Planet Imager and Characterizer: A dedicated single-mode fiber injection unit for high resolution exoplanet spectroscopy}

\author[a,b]{Jacques-Robert Delorme} %[0000-0001-8953-1008]
\author[a]{Nemanja Jovanovic} % [0000-0001-5213-6207]
\author[a]{Daniel Echeverri}
\author[a,c]{Dimitri Mawet}
\author[c]{J. Kent Wallace}
\author[c]{Randall D. Bartos}
\author[b]{Sylvain Cetre}
\author[b]{Peter Wizinowich}
\author[b]{Sam Ragland}
\author[b]{Scott Lilley}
\author[b]{Edward Wetherell}
\author[b]{Greg Doppmann}
\author[a]{Jason J. Wang}
\author[d]{Evan C. Morris}
\author[a]{Jean-Baptiste Ruffio}
\author[d]{Emily C. Martin}
\author[e]{Michael P. Fitzgerald}
\author[c]{Garreth Ruane}
\author[a]{Tobias Schofield}
\author[c]{Nick Suominen}
\author[a]{Benjamin Calvin}
\author[e]{Eric Wang}
\author[e]{Kenneth Magnone}
\author[e]{Christopher Johnson}
\author[e]{Ji Man Sohn}
\author[e]{Ronald A. L\'opez}
\author[b]{Charlotte Z. Bond}
\author[a]{Jacklyn Pezzato}
\author[a]{Jorge Llop Sayson}
\author[f]{Mark Chun}
\author[d]{Andrew J. Skemer}

\affil[a]{Department of Astronomy, California Institute of Technology, Pasadena, CA 91106, USA}

\affil[b]{W. M. Keck Observatory, 65-1120 Mamalahoa Highway., Kamuela, HI 96743, USA.}

\affil[c]{Jet Propulsion Laboratory, California Institute of Technology, Pasadena, CA 91109, USA}

\affil[d]{U.C. Santa Cruz, 1156 High Street, Santa Cruz, Ca 95064, USA.}

\affil[e]{Department of Physics \& Astronomy, University of California, Los Angeles, CA 90095, USA}

\affil[f]{Institute for Astronomy, University of Hawaii, 640 N. Aohoku Place, Hilo, HI 96720, USA.}

\cftpagenumbersoff{figure}
\cftpagenumbersoff{table}

\begin{document} 
% Adds the title and the author list
\maketitle
% Include email contact information for corresponding author
{\noindent \footnotesize\textbf{*} Corresponding author: Jacques-Robert Delorme, \linkable{jrdelorme@gmail.com} }

\begin{abstract}
The Keck Planet Imager and Characterizer (KPIC) is a purpose-built instrument to demonstrate new technological and instrumental concepts initially developed for the exoplanet direct imaging field. Located downstream of the current Keck II adaptive optic system, KPIC contains a fiber injection unit (FIU) capable of combining the high-contrast imaging capability of the adaptive optics system with the high dispersion spectroscopy capability of the current Keck high resolution infrared spectrograph (NIRSPEC). Deployed at Keck in September 2018, this instrument has already been used to acquire high resolution spectra ($R > 30,000$) of multiple targets of interest. In the near term, it will be used to spectrally characterize known directly imaged exoplanets and low-mass brown dwarf companions visible in the northern hemisphere with a spectral resolution high enough to enable spin and planetary radial velocity measurements as well as Doppler imaging of atmospheric weather phenomena. Here we present the design of the FIU, the unique calibration procedures needed to operate a single-mode fiber instrument and the system performance.
\end{abstract}

% Include a list of up to six keywords after the abstract

\section{Introduction \label{sec:Introduction}}
Since the first exoplanet detections almost three decades ago~\cite{Wolszczan1992_Nature,Mayor1995_Nature}, thousands have been detected and confirmed using various indirect and direct observing strategies. Each technique provides access to specific planet populations and allows the observer to retrieve specific parameters of the systems observed. Characterization of these planetary systems is critical to understanding their properties, formation and evolution. Photometric and spectroscopic data are particularly valuable as they give access to multiple parameters (orbital parameters, spin, temperature, atmospheric composition, cloud coverage...). 
High dispersion spectroscopy (HDS) has quickly been recognized as a powerful way to characterize exoplanet atmospheres and some strategies has been developed to spectrally characterize exoplanets detected using indirect methods~\cite{Charbonneau1999_ApJ,Moutou2001_A&A}. However, these measurements were challenging until more advanced observing strategies became available and used with stable infrared spectrographs such as Keck/NIRSPEC~\cite{McLean1995_SPIE} or VLT/CRIRES~\cite{Kaeufl2004_SPIE}. The first molecular detections in exoplanet atmospheres allowed observers to constraint orbital parameters and retrieve new information such as wind flow and molecule mixing ratios~\cite{Snellen2010_Nature}.

In parallel, the adaptive optic (AO) system and the infrared detectors of VLT/NACO~\cite{Rousset2000_SPIE,Lenzen1998_SPIE} have been used to image the first exoplanet in 2004~\cite{Chauvin2004_A&A} followed in 2008 by the detection of the first three companions to HR 8799~\cite{Marois2008_Science}, using the high contrast imaging (HCI) capabilities of both Keck and Gemini telescopes. These high HCI techniques have since been improved and many new observing strategies developed. Today most of the large telescopes have HCI capabilities, which enabled the detection of a few dozen companions. Photometric and spectroscopic data have slowly been collected on all the companions imaged using low to medium resolving power instruments (R$\sim$10--5,000) such as Keck/OSIRIS~\cite{Larkin2006_NAR}, Keck/Nirc2, LBT/LMIRcam~\cite{Skrutskie2010_SPIE}, Palomar/P1640~\cite{Oppenheimer2012_SPIE}, VLT/Sphere~\cite{beuzit2019_AAP}, Gemini/GPI ~\cite{macintosh2014_NAS} or Subaru/CHARIS~\cite{groff2016_SPIE}. 

The next step in the field is to push to higher resolving power (R$>$30,000) in order to collect spectral data where absorption and emission lines begin to be resolved by using the high dispersion coronagraphy (HDC) technique~\cite{Snellen2014_Nature, Snellen2015_A&A, Wang2017_AJ, Mawet2017_APJ,Kawahara2014_ApJS}. This technique optimally combines HCI and HDS and provides the ability to do species-by-species molecular characterization (e.g.~water, carbon monoxide, methane,...), thermal (vertical) atmospheric structure, planetary spin measurements (length of day), and potentially Doppler imaging of atmospheric (clouds) and/or surface features (continents versus oceans)~\cite{Wang2017_AJ}. As none of the current facility instruments at large telescopes offers both HCI and HDS capabilities together, several projects recently commenced to combine the HCI capabilities with the HDS capabilities of various instruments at those observatories. Among them the Fiber Injection Unit (FIU) part of the Keck Planet Imager and Characterizer (KPIC) project presented in this paper, the Rigorous Exoplanetary Atmosphere Characterization with High dispersion coronography instrument (REACH)~\cite{jovanovic2017_AO4ELT,Kotani2020_SPIE} and High-Resolution Imaging and Spectroscopy of Exoplanets (HiRISE)~\cite{vigan2018_SPIE}. The first two projects are both transitioning from commissioning to early science at the time of writing of this article and offer complimentary wavelength coverage across the near-IR (NIR) on Mauna Kea (REACH operates from Y-H and the KPIC-FIU operates in K and L bands), while HiRISE is still in the development stage. 

In this paper, we present a detailed overview of the design and performance of the dedicated FIU for the Keck II telescope. In Sect.~\ref{sec:KPIC} we briefly introduce the KPIC demonstrator in which the FIU is contained and presents its goals and the requirement who must be satisfied to reach them. In Sect.~\ref{sec:Design}, we present details of the design of the FIU, a quick description of the key components of NIRSPEC and our studies into the two key properties of the FIU: background and throughput. In Sect.~\ref{sec:Calibration}, we describe the main calibration procedures performed to prepare the FIU for science observations. Section~\ref{sec:Acquisition} presents the acquisition sequence used during the night to acquire science data and we present some example data acquired on one of our first test targets (HR 7672 B~\cite{Liu2002_ApJ}). In Sect.~\ref{sec:First_Results} we demonstrate the sorts of information that can be extract from HDC data. Finally, Sect.~\ref{sec:Conclusion} rounds out the paper with some concluding thoughts.  

\section{Keck Planet Imager and Characterizer \label{sec:KPIC}}
\begin{figure}[ht]
    \begin{center}
    \begin{tabular}{c}
    \includegraphics[width=16cm]{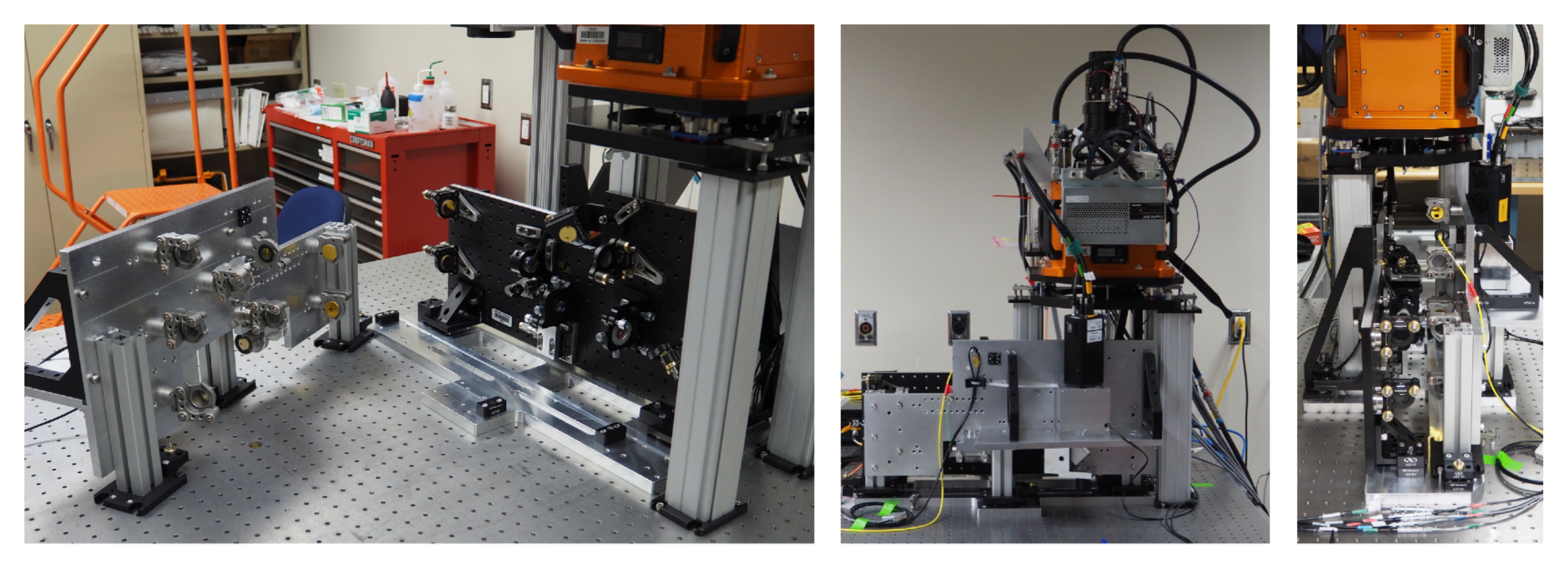}
    \end{tabular}
    \end{center}
    \caption[KPIC]{\label{fig:KPIC_Pictures} Pictures of KPIC before its deployment.}
    \end{figure}

The Keck Planet Imager and Characterizer is a demonstrator deployed at the W. M. Keck Observatory. Located downstream of the Keck II adaptive optic system~\cite{Wizinowich2000_PASP}, the goal of this platform is to demonstrate multiple new concepts to image and characterize exoplanets on-sky. In order to reduce the complexity of such a project, the project is broken into phases. Phase I, deployed at the summit of Maunakea in fall 2018, contains an infrared pyramid wavefront sensor (PyWFS)~\cite{bond2020-AOI} and the modules and components critical for the demonstration of the FIU. Figure~\ref{fig:KPIC_Pictures} presents several pictures of the KPIC optomechanics before their deployment. The vertical aluminum plate (unanodized) is the FIU plate. It supports optics commons to both the FIU and PyWFS as well as the components for the FIU. The black anodized plate is the PyWFS plate. It supports all the optics of the PyWFS. The horizontal plate located under the two previous plates is a kinematic platform to simplify co-alignment between the two vertical plates. The orange module supported by 3 struts is a SAPHIRA detector used by the PyWFS.

%-------------
%-------------

The second phase, planned to deploy in late 2021 will contain extra modules: an atmospheric dispersion compensator, phase induced amplitude apodization optics, a high order deformable mirror and coronagraphs, whose goal will be to improve the overall throughput for the planet light, while reducing star light leakage to reduce overall integration times\cite{Jovanovic2019_SPIE, pezzato2019-SKP, jovanovic2020_SPIE}. In this paper, we present a detailed overview of the first phase FIU and its performance. 

\subsection{FIU phase I goals and requirements \label{sec:Goals}}

The first version of the FIU presented in this paper has been designed as a prototype. The design has been driven by the space available in the Keck II AO bench, the existing science instruments available at Keck (i.e. NIRC2 and NIRSPEC) and the science goals. Its main goal was to demonstrate the concept by acquiring high resolution spectral data on a variety of targets. We wanted the instrument to be able to observe of all known substellar companions accessible to Keck, which sets most of the requirements as shown in Table~\ref{tab:Requirements}.

To achieve this goal, the target or its host star must be observable by the Keck II AO system. Although the FIU can be used in combination with the visible Shack Hartmann wavefront sensor, we mainly use it with the IR PyWFS. Since its deployment, this wavefront sensor, which operates in H band, has been consistently used to observe targets up to $12^{th}$ magnitude. The Strehl ratio measured in K band is ~65\% in 0.5 arcsec seeing and ~45\% in 1 arcsec seeing when the H magnitude of the star observed is $\leq 6$. Above this magnitude performance progressively decreases. The performance of this wavefront sensor has been described in detail in Ref.~\citenum{bond2020-AOI}. It is currently not possible to close the AO loop on a system with two or more components if they have a similar H magnitude and a separation between 0.5 and 2 arc-seconds. This issue will be addressed when the second phase will be deployed. Because the PyWFS is using most of the H band light and the targets of interest are brighter in the K and L bands, the first version of the FIU was designed to collect spectral data in those photometric bands. 

Once the AO loop is closed, the main challenge is to precisely align the object of interest with a single mode fiber (SMF) with a precision of less than one fifth of a $\lambda/D$ ($\leq$ 10 mas in K band) to achieve and maintain high coupling. To align the object of interest with a SMF, we image the system on a detector. Usually not visible on this detector, the object of interest is aligned with a SMF by tracking the position of its host star at a relatively low frequency ($<1$Hz). This camera is sensitive to both J and H bands. However, we permanently installed a H band filter in front of it. In the second phase of this project we will installed it in a filter wheel with multiple filter options.In order to observe all known substellar companions accessible to Keck, we must be able to track the position of object with a H-mag up to 12 and have a capture range of $\pm 2.5$ arcsec.

The use of SMFs makes the observation with such a system more challenging than observations with conventional AO-fed slit spectrographs such as Keck/NIRSPEC or VLT/CRIRES. However, they provide multiple benefits. Specifically, the SMF has a small solid angle on sky minimizing thermal background leaking into the spectrograph, which is critical in K and L band. They can also help filter out unwanted speckle (star light)~\cite{Mawet2017_SPIE} and they make future wavefront control relatively simple in that only a few spatial frequencies need to be addressed~\cite{Llop-Sayson2020_SPIE}, which allows the bandwidth to be increased with these algorithms~\cite{Coker2019_SPIE}. Finally, they allow the use of compact fiber-fed spectrographs that can be located of the telescope in a more stable environment. If fed by an SMF, the position and shape of the beam inside a fiber-fed spectrograph is agnostic of input conditions and greatly simplifies the spectral calibration by making the line spread function, a Gaussian profile in this case, which is extremely stable.

In order to reduce the cost and the time needed for this project, we designed the FIU as an interface to connect the AO bench to NIRSPEC. Because NIRSPEC was not designed as a fiber-fed spectrograph, it constrained the design in multiple ways (shape of the cold stop, injection outside of the cryostat, size and shape of the slits available, etc) described later in this paper.NIRSPEC sets the resolving power to 35,000 if we maintain a sampling of at least 2 pixels at all wavelengths across the bands observed (K or L band). The data collected are compared to models by computing the cross correlation function (CCF). To be usable, we must be able to detect the spectral signature of the object of interest with a signal-to-noise ratio (S/N) greater than 3 after two hours of integration. The table~\ref{tab:Requirements} summarizes all the baseline requirements for the KPIC FIU.

%-------------
    \begin{table}[ht]
    \begin{center}
    \begin{tabular}{ |c|c|c|c|  }
    \hline
    \hline
    \multicolumn{3}{|c|}{\textbf{Tracking and acquisition}} \\
    \hline
    \hline
    \textbf{Specifications} & \textbf{Value} & \textbf{Notes}\\
    \hline
    \hline
    Scale of optical fibers & $\sim\lambda/D$ & SMF optimally matched to PSF \\
    \hline
    Capture range & $\pm2.5$~arcseconds & Covers most high contrast targets\\
    \hline
    Alignment accuracy & $<0.2 \lambda/D$ & ---\\
    \hline
    Tracking speed & $<1$ Hz & Tracking, not tip/tilt correction\\
    \hline
    Tracking stability & $<0.2\lambda/D$ RMS & ---\\
    \hline
    Tracking bands & J or H band & ---\\
    \hline
    Limiting magnitude & $>12$ in H-mag & PyWFS limitation\\
    \hline
    \hline 
    \multicolumn{3}{|c|}{\textbf{Spectroscopy}} \\
    \hline
    \hline
    \textbf{Specifications} & \textbf{Value} & \textbf{Notes}\\
    \hline
    \hline
    Spectral Range & K \& L band & Not simultaneously \\
    \hline
    Resolving Power & $\geq 30,000$ & NIRSPEC limitation\\
    \hline
    Sampling & $\geq 2$ pixels & At all wavelengths \\
    \hline
    \hline
    \multicolumn{3}{|c|}{\textbf{Overall performance}} \\
    \hline
    \hline
    \textbf{Specifications} & \textbf{Value} & \textbf{Notes}\\
    \hline
    \hline
    Total Peak Efficiency & $\ge2\%$ & From top of atmosphere\\
    \hline
    Pt sources sensitivity & $\geq 16$ K-mag & S/N $\ge$ 3 in CCF in $\approx$ 2h.\\ 
    \hline
    \end{tabular}
    \end{center}
    \caption[Requirements]{\label{tab:Requirements} FIU Phase I requirements.}
    \end{table}
    
%-------------

\section{Fiber injection unit design \label{sec:Design}}
The FIU deployed at Keck is based on three modules: an injection module, a bundle of fibers and an extraction module. This section describes these modules, the NIRSPEC instrument used to acquire the data and the expected performances of the overall instrument. The interaction between these components and the way we are using them to achieve our goals is described in the following sections of this paper. 

\subsection{Injection module \label{sec:FIU}}
\begin{figure}[ht]
    \begin{center}
    \begin{tabular}{c}
    \includegraphics[width=16cm]{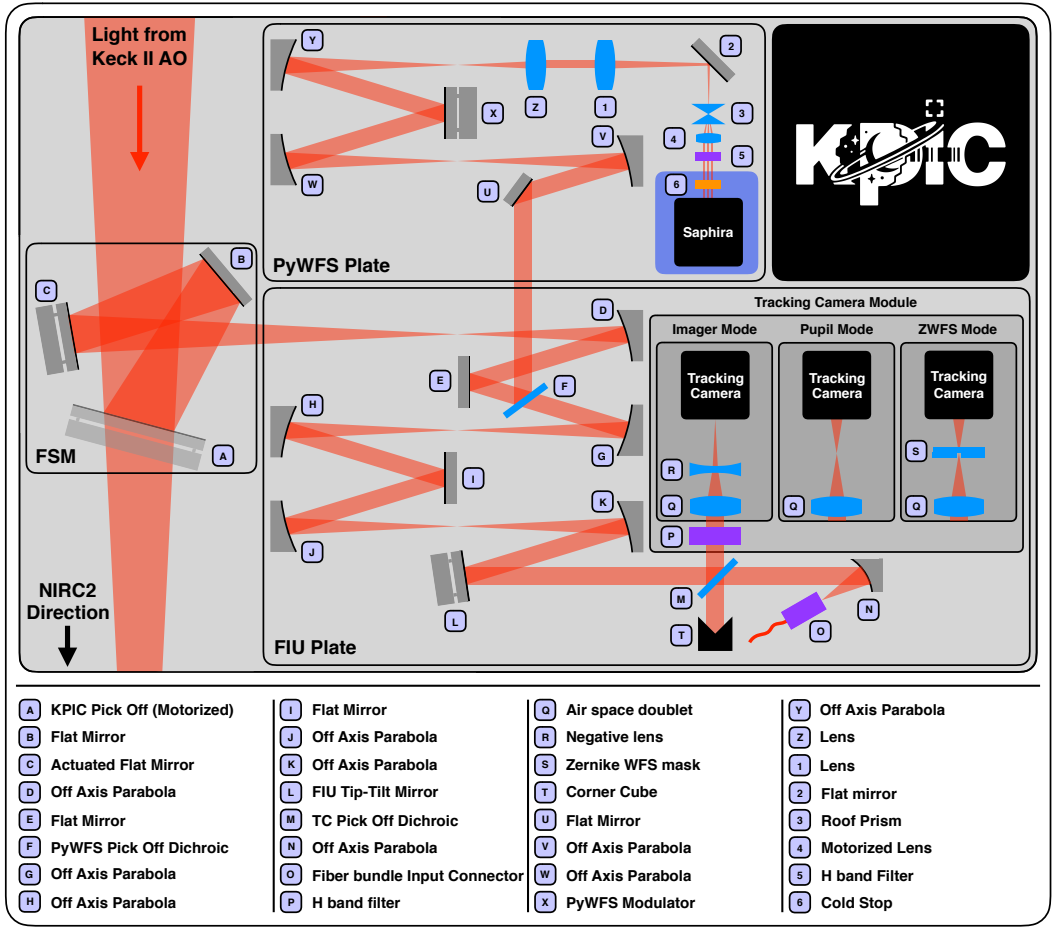}
    \end{tabular}
    \end{center}
    \caption[FIU]{\label{fig:KPIC_Diagram} Diagram of the KPIC optical layout. The scale and the orientation of the optics are not correct. The FIU and PyWFS plates as well as the Saphira detector are visible in the first figure of this paper.}
    \end{figure}

Located downstream of the Keck II AO system the goal of the injection module is to inject the light of faint companions (exoplanets, brown dwarf, etc) into one of the single-mode fibers (SMFs) of a bundle described in Sect.~\ref{sec:Bundle}. Figure~\ref{fig:KPIC_Diagram} shows a schematic of the KPIC optical layout. This diagram is not to scale and the orientation of the optics is not correct. In this paper, we only describe the optical layout of the FIU.

%-------------
%-------------

To direct the light to the FIU, the converging beam coming from the AO bench is reflected by the three optics constituting a field steering mirror module (FSM - See Fig.~\ref{fig:KPIC_Diagram}). The first optic of this FSM (A) is the KPIC pick off. It consists of an actuated dichroic and a static flat mirror installed on a translation stage. When neither optic is in the beam, all the light goes to the facility infrared imager NIRC2. If the dichroic is in the beam, 90\% of the J and H band light goes to KPIC while the rest goes to NIRC2 . If the flat mirror is in the beam, all the light goes to KPIC.

After the FSM, the converging beam is directed to the FIU plate. This plate is visible in the images presented in Fig.~\ref{fig:KPIC_Pictures} (unanodized), and supports all the components of the injection module. After going through two optical relays (D-J in Fig.~\ref{fig:KPIC_Diagram}), comprised of simple reflective optics in phase I, the light is collimated by a 168.31 mm focal length off axis parabola (OAP, K) made of Zerodur and coated with protected gold (\textit{Nu-Tek}). The two optical relays are based on four identical OAPs to the one used for collimation (\textit{Nu-Tek} - D, G, K, J) and two flat mirrors (\textit{Newport} - E and I). All these optics are made of Zerodur and coated with protected gold. The F\# of the input and output focal plane are the same as the F\# of the telescope (13.66). The diameter of the collimated beams in these relays is 12.32 mm. In the first optical relay, the static pick off dichroic of the PWS (\textit{Asahi Spectra} - F) reflects 90\% of the J and H band light in the direction of the sensor while the rest of the light is transmitted (K and L band). These two optical relays mostly empty during the first phase of the project will be populated in the phase  two~\cite{pezzato2019-SKP,jovanovic2020_SPIE}.

After the two relays, a tip-tilt mirror (TTM) is situated in the pupil plane of the instrument for aligning the target with the fiber as well as fine tuning the pointing onto the fibers (L). The TTM (\textit{Physik Instrumente}, S-330.8SL) is a piezo mechanism, which offers a field steering of $\pm$ 2.4 arc-second in two axes. An off-the-shelf protected gold flat mirror (\textit{Thorlabs}, 19~mm diameter) is attached to the platform using a custom made, flexure-based mirror holder. Made of Invar and titanium, this mirror holder has been designed to minimize the forces applied to the flat mirror in order to minimize distortions and to be as light as possible in order to optimize the response of the overall TTM module. 

A custom dichroic (\textit{Asahi Spectra}, M) is used to split light and direct it to the camera for tracking purposes while allowing science light to pass through to the injection unit. It reflect 90\% of the J and H band light while longer wavelength light ($\geq$K and L) is transmitted.

The reflected path towards the camera contains a H band filter (\textit{Asahi Spectra}, P), an air-spaced achromatic doublet (\textit{Thorlabs}, Q), a plano-concave lens (\textit{Edmund Optics}, R) and a low noise InGaAs detector (\textit{First Light Imaging}, Cred-2~\cite{Gibson2019_SPIE}), which offers a 4.125 $\times$ 5.125 arc-second field-of-view (FOV, 512 x 620 pixels, with an $\approx$ 8.06 mas/pixel sampling). When KPIC is used as a FIU, this detector is used to track the position of the science targets. For this reason, we refer to it as the ``tracking camera". A stage can be used to translate the plano-concave lens in/out of the beam. When out of the beam, a pupil image of the Keck primary is formed. When in the beam, a focal plane image is formed. The FOV of this detector has been carefully selected to be able to observe the targets of interest selected for this project. It limits the maximum separation between the objects of the systems observed. However, this FOV allows us to observe most of the exoplanets already imaged and visible from Maunakea.
The K and L band light is next incident on the injection optic (N): a 35 mm focal length OAP (\textit{Nu-Tek}). Made of Zerodur and coated with protected gold, this optic is the last of the FIU module before the fiber bundle (O) described in the following section. The bundle is carefully positioned at the focus (F\# = 2.8) and on the optical axis of the OAP. The effective focal length of the OAP is the one which maximizes the injection into the fiber for both K and L bands.

To aid in aligning the incoming beam with the bundle, a corner cube (T) is located beneath the dichroic. Its role will be explained in the calibration procedures of the system (see Sect.~\ref{sec:Cals_ScF_Position} ).\\

\subsection{Fiber bundle \label{sec:Bundle}}
\begin{figure}[ht]
    \begin{center}
    \begin{tabular}{c}
    \includegraphics[width=16cm]{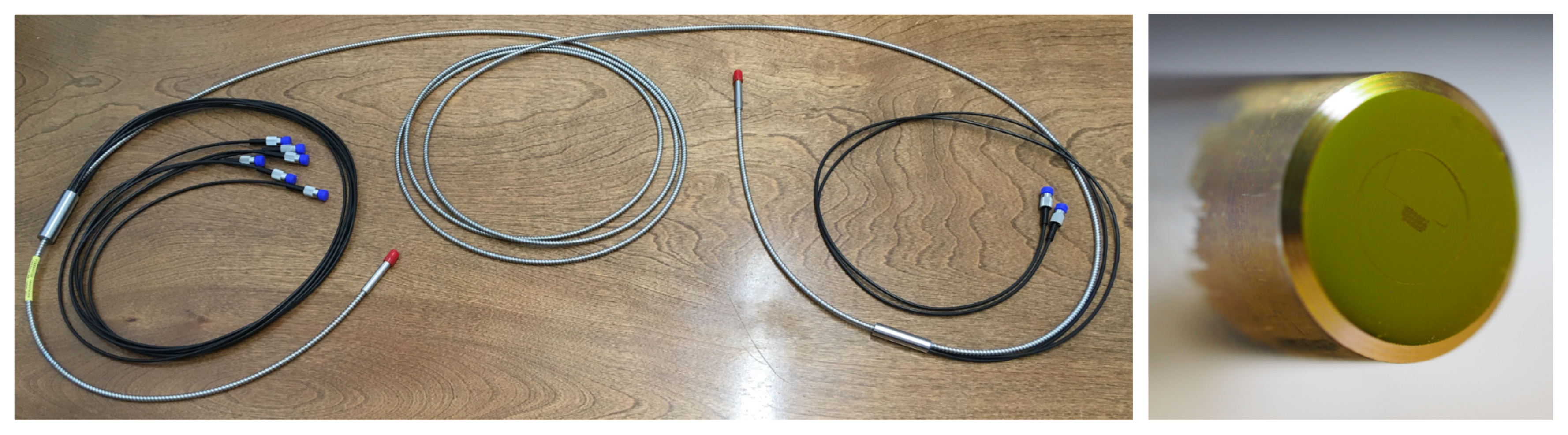}
    \end{tabular}
    \end{center}
    \caption[Bundle diagram]{\label{fig:bundle_diagram} Diagram the first KPIC-FIU fiber bundle.}
    \end{figure}

The fiber bundle has several key roles: 1) it is used to route the science light to the spectrograph and 2) it has peripheral fibers that light can be reverse injected into to aid with alignment optimization. The layout of the bundle used in KPIC is shown in Fig.~\ref{fig:bundle_diagram}. On the input end, the fiber bundle is connected to the injection module (see Sect.~\ref{sec:FIU}) located inside the Keck AO bench. On the output side, the fiber bundle is connected to the extraction module (see Sect.~\ref{sec:FEU}) located in the calibration unit of NIRSPEC~\cite{McLean1998_SPIE} a facility class high spectral resolution spectrograph of the W. M. Keck Observatory.

%-------------
%-------------

Manufactured by \textit{Fiber Guide Industries} this critical component of the FIU is based on ZBLAN 6.5/125 fluoride fibers (\textit{Le Verre Fluor\'e}) and SMF-28-ULTRA silica fibers (\textit{Corning}). The bundle contains five ZBLAN 6.5/125 science fibers going from the input to the output connectors (note, this indicates the fibers have a $6.5~\mu$m diameter core and $125~\mu$m cladding). These are indicated in red on the diagram. This part of the bundle is 5 m long, a length necessary to connect the injection and extraction modules. The science fibers are surrounded by six SMF-28-ULTRA calibration fibers at the input connector end and two ZBLAN 6.5/125 calibration fibers on the output connector end, represented in blue and orange in the figure respectively. These alignment fibers are contained within the furcation tubing of the main bundle for about $50$~cm before they are broken out into individual fibers with FC/PC connectors. 
The input and output connectors also contain dummy fibers which act as scaffolding to locate the science and calibration fibers during manufacturing. Represented in black in the diagram, these fibers are segments of SMF-28-ULTRA terminated within the package.

The custom input and output connectors were manufactured using two different approaches compatible with the specific needs of each connector. The input connector contains three layers of fibers where the position of each fiber is not critical. In this connector, the fibers are butted cladding-to-cladding and their position is maintained by an outer metallic structure. The cladding-to-cladding construction maintains the core separation of two adjacent fibers to $\approx 125\:\mu$m which corresponds to $\approx 800$ mas in K band. 

The output connector contains only one layer of fibers but these fibers must be aligned with a high precision to make sure they pass through the mechanical slit of NIRSPEC downstream. For this reason, the fibers are supported using a V-groove on this side. The pitch of the V-groove used set the core-to-core separation between two adjacent fibers to $127\:\mu$m. On the input side, four of the input calibration fibers, labeled $C_{1}$, $C_{2}$, $C_{4}$ and $C_{6}$ in the diagram, are connected to a 1550~nm laser source. The two other input calibration fibers ($C_{3}$ and $C_{5}$) were connected to two high speed InGaAs photo-detectors. On the output side, the two calibration fibers ($C_{A}$ and $C_{B}$) were connected to two broadband infrared light sources (tungsten lamps, Thorlabs, SLS202L). In section \ref{sec:Calibration} we describe how and when these elements are used.

\begin{figure}[ht]
    \begin{center}
    \begin{tabular}{c}
    \includegraphics[width=16cm]{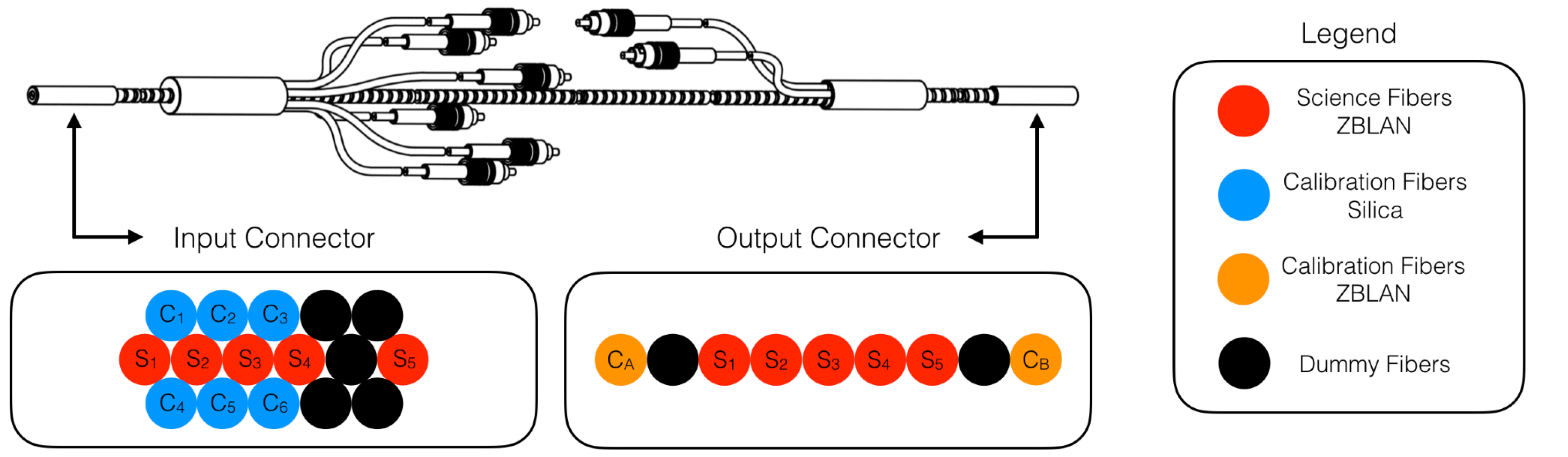}    
    \end{tabular}
    \end{center}
    \caption[Bundle pictures]{\label{fig:bundle_pictures} Left: Picture of the bundle. Right: Picture of the input connector of the bundle.}
    \end{figure}

Figure~\ref{fig:bundle_pictures} shows a picture of the fiber bundle currently used by KPIC (left) and a picture of the input connector (right). For reference, the diameter of the connector is 7 mm. The gold coating visible on the input connector is an anti-reflection (AR) coating to optimize the throughput of the bundle in K and L bands, and is used at the output as well. The overall throughput of the bundle in K band was measured in the laboratory to be $95.3\%$ in K band. According to the manufacturer, the attenuation of the ZBLAN fiber varied between 2.6 and 5.3 dB/km across the K-band which corresponds to a throughput $\geq 99.4\%$ for 5 meters of fiber. The rest of the throughput loss is due to the Fresnel reflections which occur at both ends of the bundle due to imperfect AR coatings. The L-band throughput has not been measured because of a lack of equipment in those bands. However, some preliminary observations in L-band indicate that the throughput of the bundle in this band is close to expectation.\\

%-------------
%-------------

The minimum number of fibers needed to acquire HDC data is three: one for the target of interest, one for its host star and one for the sky background. The final number of science fibers was chosen as a trade off between the price of the bundle and various technical constraints. For instance, the shape and size of the NIRSPEC slit limits the number of fibers on the output connector and imposes a linear arrangement of fibers. We decided to use a design with five science fibers to cover most science cases and to have redundancy in case one of the science fibers was damaged. Regarding the calibration fibers, the bundle has been designed to be functional even if one of the calibration fibers of each set is damaged (one fiber connected to the laser, one to the photo-detector and one to the broadband light source). The position of the fibers on the input connector was not critical. We required the arrangement of fiber relatively to be compact in order to image most or all of them on the tracking camera if needed. To keep the system simple and efficient, we chose to maintain a coarse separation between the fibers. The minimum separation between two fibers is $\approx 800$ mas and the maximum is $\approx 4000$ mas. Because the star is usually brighter than the companion by several magnitudes, we can collect stellar light even if the star is not aligned with a fiber (light from the speckle field). We can adjust the amount of stellar light injected into a science fiber using the rotator of the telescope by bringing the star closer to a science fiber or moving it away. If the host star cannot be brought close enough to a science fiber when data are collected on the companion, we can collect spectral data on the host star before or/and after by aligning it with one of the science fibers. Finally, because the separation between the first and last science fiber is $\approx 4$ arcsec, one of the fibers will always be far enough from the star to acquire background data.

\subsection{Extraction module \label{sec:FEU}}
\begin{figure}[ht]
    \begin{center}
    \begin{tabular}{c}
    \includegraphics[width=16cm]{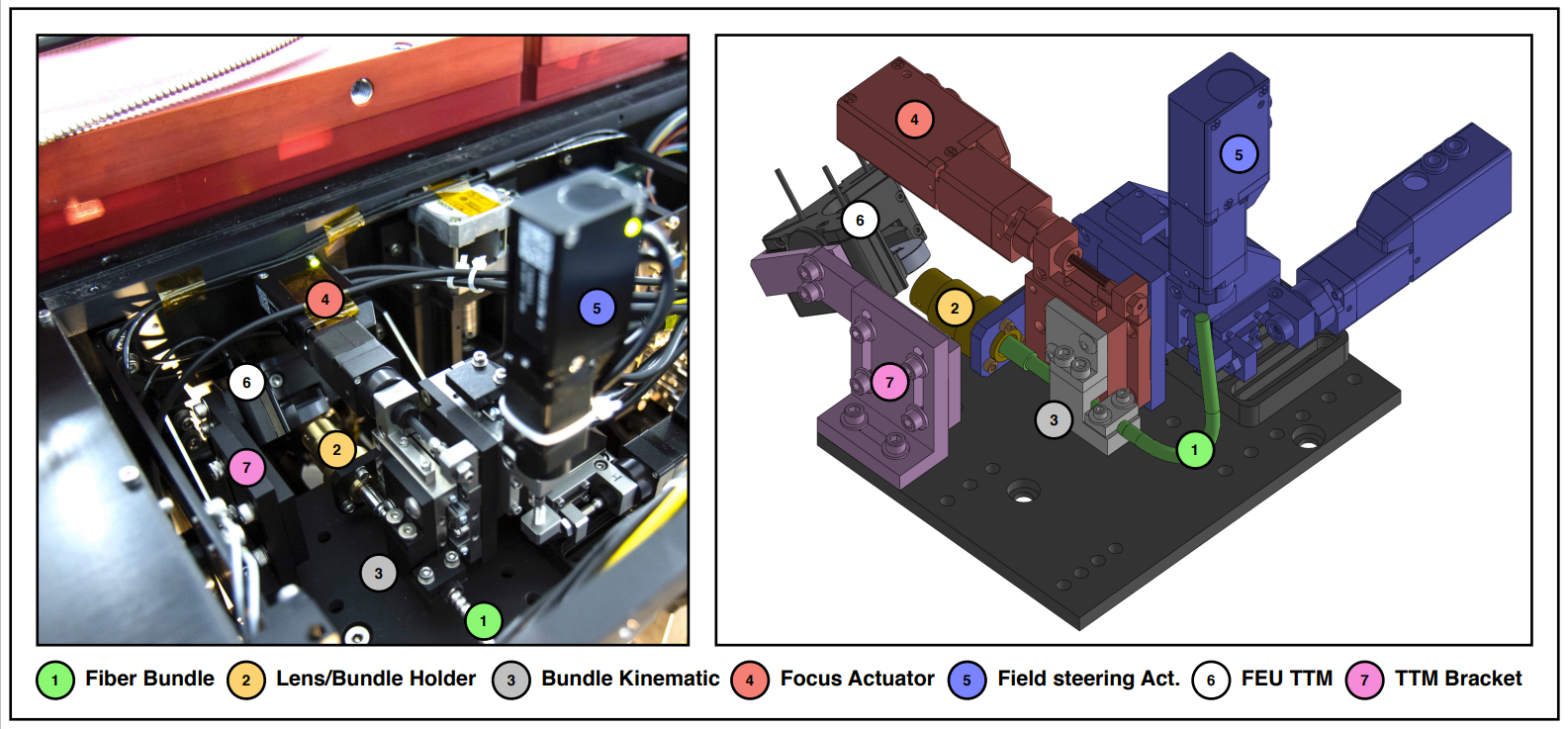}
    \end{tabular}
    \end{center}
    \caption[FEU]{\label{fig:FEU} Left: picture of the FEU installed in NIRSPEC's calibration unit. Right: a CAD model view of the FEU.}
    \end{figure}

The extraction module, known as the fiber extraction unit (FEU) is the interface between the fiber bundle and NIRSPEC. Its goal is to reshape the light coming from the different fibers of the bundle and optimally inject this light into the high-resolution spectrograph, which was originally designed to work in the seeing limit. The left panel of Figure~\ref{fig:FEU} shows a picture of the FEU installed in the calibration unit of NIRSPEC, and the right panel is the computer-aided design (CAD) model.

%-------------
%-------------

A custom made brass part visible in the right panel of Fig.~\ref{fig:FEU} (labeled 2) guides the output connector of the bundle (labeled 1) and aids in aligning it with an air-spaced doublet collimating lens mounted in the same part. This part has been made of brass mainly because this material is relatively soft and easy to machine. On the doublet side, we built a custom  barrel to match the diameter of the lenses and to adjust the distance between the lenses if needed. For the bundle, we chose the material of the barrel to be softer than the material of the bundle (aluminum) to minimize the risk of galling when the bundle slides in the barrel. The doublet is composed of a germanium and silicon lens (\textit{Rocky Mountain Instrument}). The effective focal length of this doublet is 10.84 mm and the throughput in K and L bands was measured to be $>92\%$. The numerical aperture of the science fiber has been measured in laboratory to be 0.175 at 2 $\mu$m. After the air-spaced doublet, the $1/e^{2}$ diameter of the collimated Gaussian beam is 3.8 mm. Close to its output connector, the stainless steel mono-coil jacket of the bundle is attached to a motorized translation stage (labeled 4) using a custom made kinematic interface (labeled 3). The translation stage is used to move the bundle along the optical axis guided by the brass part, which corresponds to the focus axis of the bundle/lens system. The brass piece prevents the bundle from moving in the axes perpendicular to the optical axis and constrains the pitch and yaw of the output connector while the bundle is being focused. The rotation of the bundle about the optical axis is constrained by a clamp and kinematic interface. This simple kinematic interface combined with the brass holder have proved reliable when it comes to returning the fiber bundle to the same position and orientation each time it has been connected. This assembly is mounted on two motorized translation stages, labeled 5, used as a pupil steering mechanism to move the fiber bundle and the air spaced doublet in the plane perpendicular to the optical axis. This mechanism is used each time the bundle is connected to align the beam coming from the bundle to the NIRSPEC field stop located into the cryostat. The next and last optical element of the fiber extraction unit is a TTM (labeled 6) held by a custom made bracket (labeled 7). Based on a \textit{Newport} CONEX-AG-M100D, this element reflects the light in the direction of NIRSPEC calibration unit. This TTM is located in a pupil plane and is used to align the images of fibers with the slit of the spectrograph.

\subsection{NIRSPEC \label{sec:NIRSPEC}}
\begin{figure}[ht]
    \begin{center}
    \begin{tabular}{c}
    \includegraphics[width=16cm]{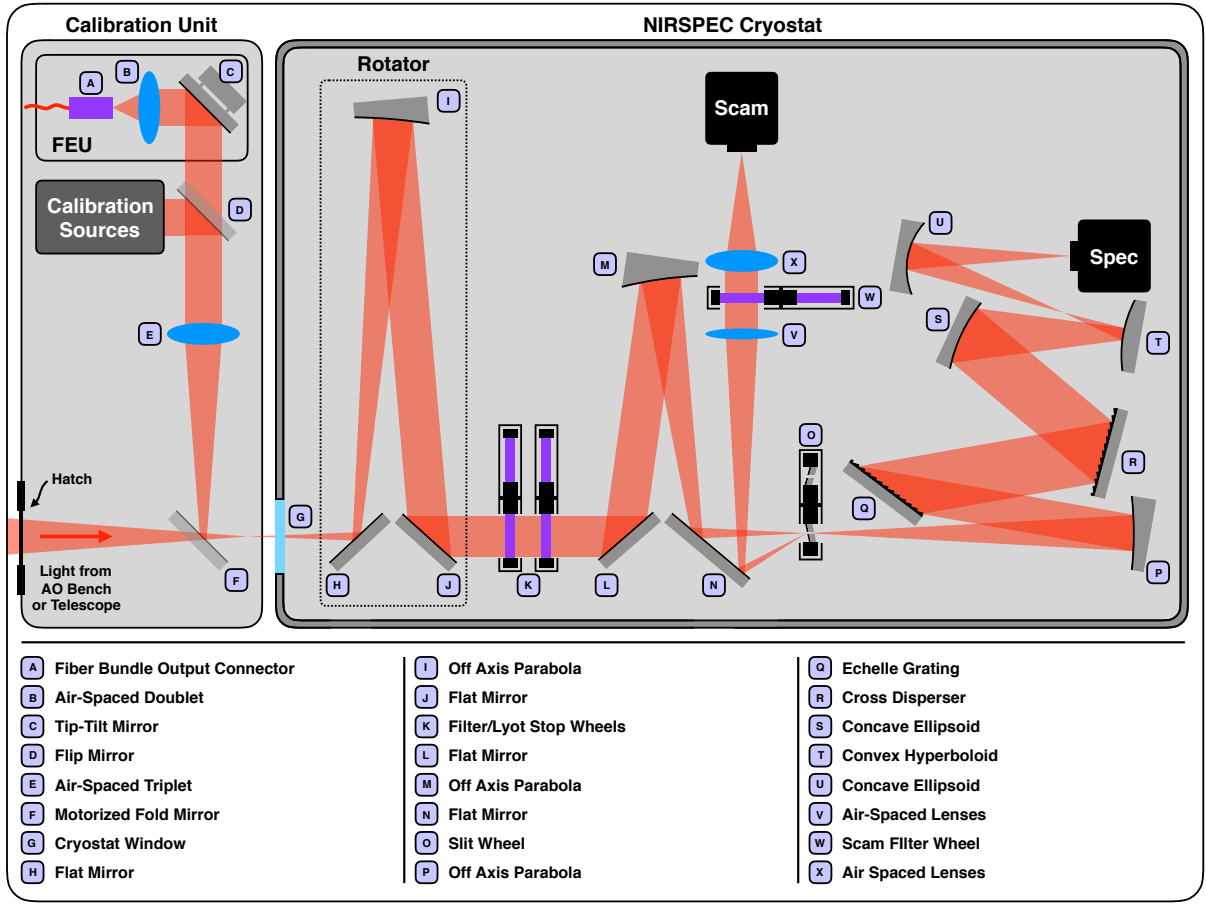}
    \end{tabular}
    \end{center}
    \caption[FEU]{\label{fig:NIRSPEC} Diagram of the NIRSPEC optical layout with the FEU installed in its calibration unit. This diagram is not to scale and does not show the optics in the correct orientation.}
    \end{figure}

Figure~\ref{fig:NIRSPEC} presents a diagram of the optical layout of NIRSPEC (not to scale). For more information regarding the NIRSPEC optical layout please consult: Ref.~\citenum{McLean1998_SPIE},~\citenum{Robichaud1998_SPIE} and~\citenum{Martin2018_SPIE}. NIRSPEC's calibration unit located to the left side of this diagram contains the FEU described in the previous section.  

%-------------
%-------------  

A fold mirror (labeled D) installed in the calibration unit of NIRSPEC was motorized using a flip mechanism (\textit{OWIS},KSHM 65-LI-MDS). It is used to select the input of the calibration unit. When in the beam, the mirror directs light coming from the calibration sources of NIRSPEC's integrating sphere (and an air-spaced collimating triplet) to the slit in order to calibrate the spectrograph. When out of the beam, the light from the FEU is directed to NIRSPEC. After this optical element, the light is focused by an air-spaced triplet (E - EFL = 152.4 mm) and reflected by a motorized fold mirror (F) before it enters into the NIRSPEC cryostat. When KPIC is in use, this fold mirror is in the beam. After this mirror, the beam goes through the uncoated CaF$_{2}$ window of the NIRSPEC cryostat before being reflected by the three mirrors of the rotator. The second optic of this rotator (I) is an OAP (EFL = 400 mm). After the rotator the collimated Gaussian beam ($1/e^{2}$ diameter of 10.0 mm) passes through two filter wheels (K) before being focused by an OAP (M) onto the selected slit located in a slit wheel (O). All NIRSPEC slits are reflective. The light passing through the slit is dispersed and imaged on the science detector named `Spec' (\textit{Teledyne}, science grade Hawaii-2RG detector). The light reflected by the slit is imaged by a slit-viewing camera named `Scam' (\textit{Teledyne}, engineering grade Hawaii-2RG detector). 

\begin{figure}[ht]
    \begin{center}
    \begin{tabular}{c}
    \includegraphics[width=12cm]{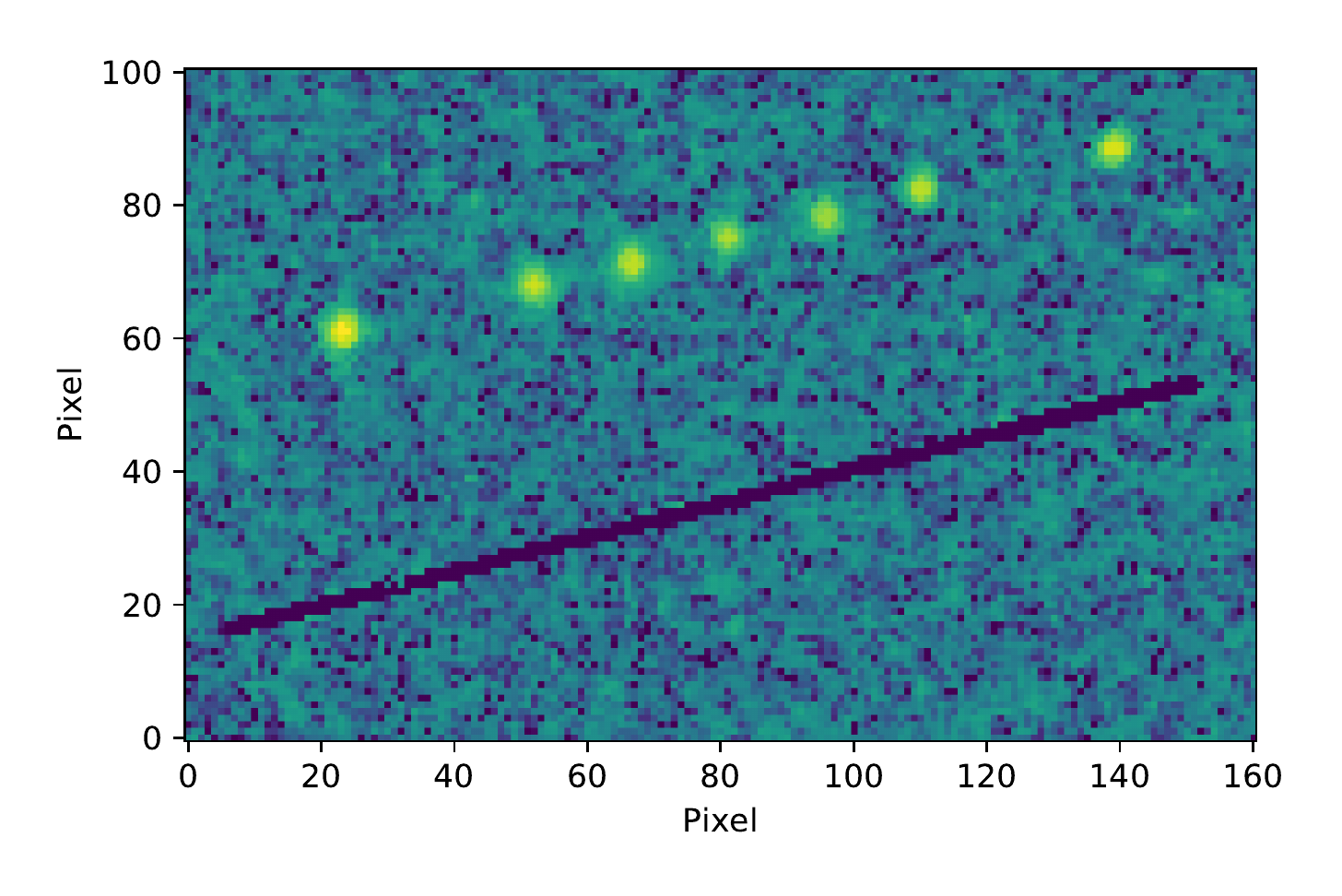}
    \end{tabular}
    \end{center}
    \caption[Scan]{\label{fig:Scam_Patchwork} A composite image with all fibers illuminated acquired with the slit-viewing camera. In this case, the seven PSFs are offset from the slit. The seven PSFs visible on this image are respectively associated to (from left to right): C$_{A}$, S$_{1}$, S$_{2}$, S$_{3}$, S$_{4}$, S$_{5}$ and C$_{B}$. The slit (dark blue) used in this case was the 0.0407"x1.13".}
    \end{figure}

Figure \ref{fig:Scam_Patchwork} presents a combination of images acquired with the slit-viewing camera. This image has been obtained by combining images acquired with light injected into each fiber of the output connector of the bundle (see Sect.~\ref{sec:Bundle}). This image presents the case where the PSFs associated with the fibers has been offset from the slit. This mode is mainly used during the calibration of the instrument. Indeed, unlike the science detector, the light coming from the fiber is not dispersed in the focal plane of the slit-viewing camera, which allows quick flux measurements. To align the PSFs and the slit, we use the FEU TTM (C). If the fibers contained in the bundle are not properly oriented with respect to the slit, we use the internal NIRSPEC rotator to make small adjustments.

%-------------
%-------------

\begin{figure}[ht]
    \begin{center}
    \begin{tabular}{c}
    \includegraphics[width=16cm]{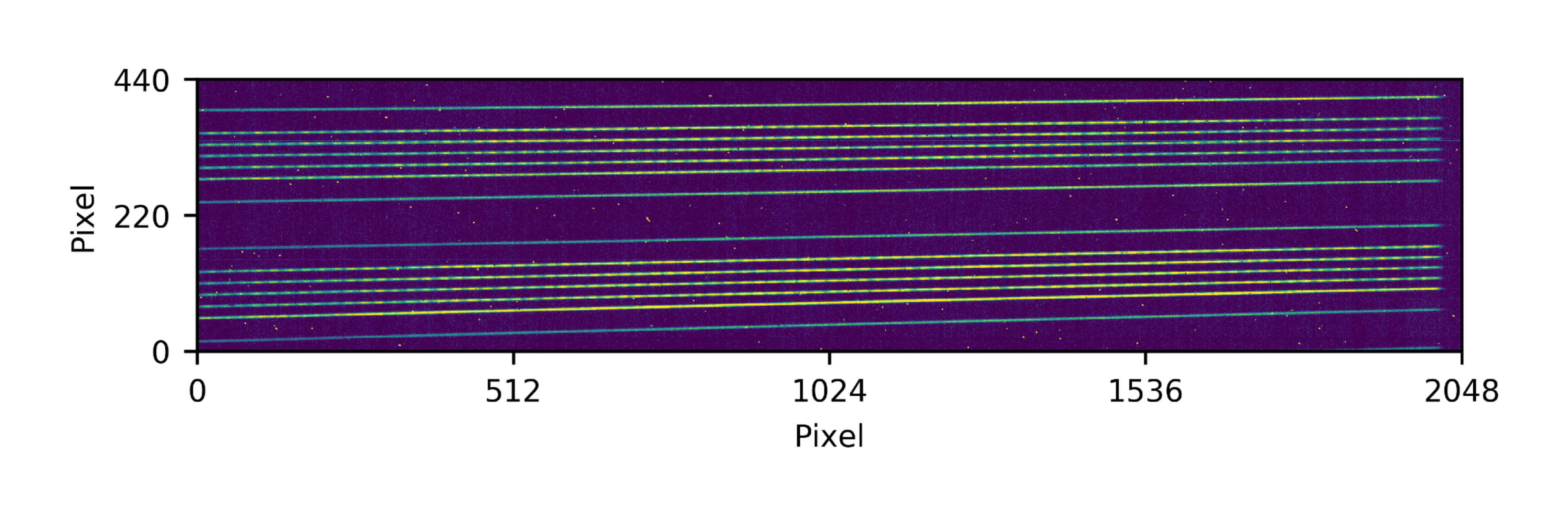}    
    \end{tabular}
    \end{center}
    \caption[Scan]{\label{fig:Spec_Patchwork} Two of the nine orders visible on the science detector when configured for K band observations. Each order of this patchwork of images contained seven spectra. The upper and lower spectra of each order are associated with the two calibration fibers contained in the output connector of the bundle while the five other spectra are associated with the science fibers. This image is a composite image with all fibers illuminated}.
    \end{figure}

Figure~\ref{fig:Spec_Patchwork} presents two of the nine orders visible on the science detector when configured for K band observations. This image is also a combination of images obtained by combining images acquired with light injected into each fiber of the output connector of the bundle. Each order contains seven spectra. The five central spectra are associated with the five science fibers of the bundle while the two lateral spectra are associated with the two calibration fibers. Depending on the slit selected (short or long), the light from the calibration fibers is reflected in the direction of Scam or transmitted to Spec. The separation between the spectra is fixed ($\approx 19$ pixels) and set by the separation between the fibers in the output connector of the bundle. The width of the trace can be adjusted by moving the bundle along the optical axis using the focus actuator (labeled 4 in Fig.~\ref{fig:FEU}). We adjust the position of this actuator during a calibration procedure of the FEU (see Sect.~\ref{sec:Cals_FEU}) to properly sampled the data.

%-------------
%-------------

\subsection{Expected performance \label{sec:Performances}}
Because the FIU module of KPIC aims to spectrally characterize faint objects with a high resolution ($R>$30,000), the background seen by the science instrument and the overall throughput are the key properties that need to be understood. In this section we present our estimations and measurements of these parameters in K band. Although, the phase I version of the instrument has been designed to allow science observations in K and L bands, we have mainly focused our work on the K band to date. We will start to work in L bands once most of the other challenges of this project will be overcome.

\subsubsection{Background \label{sec:Background}}
For a fiber-fed spectrograph, the background seen by the science instrument can be decomposed into four components: the background injected into the science fibers of the bundle (sky background and upstream optical elements), the emissivity of the fiber itself, the background emitted between the fiber bundle and the cold stop of NIRSPEC and the background emitted in the cryostat. 

The first component of the background is relatively small. SMFs are extremely efficient at rejecting upstream background from sky, telescope and FIU because of their relatively small solid angle projected on sky (45 mas$^{2}$ in K band). For reference, the average sky brightness in K band at zenith during dark times and during a clear night at the summit of Maunakea is equal to 12.6 mag/arcsec$^{2}$  (\hyperlink{http://www.cfht.hawaii.edu/Instruments/ObservatoryManual/CFHT_ObservatoryManual_\%28Sec_2\%29.html}{http://www.cfht.hawaii.edu/}).
The second component of the background is the emissivity of the fiber itself. We have not quantified this component yet. However, KPIC uses a relatively short (5 m) SMF with very high transmission and with a very small core diameter ($6.5~\mu$m). We can compute the background emitted based on the energy absorbed by the material of the fiber ($\leq 0.6\%$ in K band). However, it is difficult to estimate the fraction of this background transmitted to the FEU side of the bundle.

The third component of the background is the sum of the emissivities of all the warm optics used to project the light from the fiber onto the NIRSPEC slit (labeled A-G in the Fig.~\ref{fig:NIRSPEC}). To reduce this component of the background, we use a cold stop, named "FEU stop", located in both filter wheels of NIRSPEC (labeled K on the Fig.~\ref{fig:NIRSPEC}). These stops have been installed in both wheels in order to be able to use them with all the NIRSPEC filters available. The stops consist of a hole in a metallic plate, which has been black painted. The hole diameter has been optimized to maximize the signal-to-noise ratio (SNR) seen by the science detector. The radius of the cold stop which maximizes the SNR `$R_\text{max}$' is:  
$$ R_\text{max} \approx 1.5852 \sigma \approx 0.6731 \text{FWHM} $$
where $\sigma$ and FWHM (Full Width Half Maximum) characterize the Gaussian profile’s width\footnote{\url{https://wise2.ipac.caltech.edu/staff/fmasci/GaussApRadius.pdf}}. By using the ratio between the $1/e^{2}$ diameter and the FWHM  ($2\sqrt{2ln(2)} \approx 1.7 $), we can compute the optimal diameter for the cold stop. Because the mode field diameter of the science fibers as well as the effective focal length of the air-spaced doublet and triplet (labeled B and E on Fig.~\ref{fig:NIRSPEC}) are wavelength dependent, the optimal diameter is also wavelength dependent. Moreover, the diameter of the beam at the location of the stop can be affected by the defocus applied to the bundle to optimize the size of the PSF on the slit (see Sect.~\ref{sec:Cals_NIRSPEC}). At 2~$\mu$m, we computed the optimal diameter of the stop to be $\approx$7.9 mm. We decided to use a slightly larger cold stops (8.1 mm diameter) because the optical elements of the NIRSPEC calibration unit were never calibrated (as it wasn't needed) and because the K and L bands go beyond 2 $\mu$m. Moreover, the impact of increasing the thermal background sightly is not as severe as restricting the light from the companion.
 
These stops are used in addition to a blocking filter, which suppresses the science and background light outside of K band to prevent order overlap. Several filters are available in each filter wheel. Thus far, we have mainly observed in K band (1.9 to 2.5 $\mu$m). In this case, we are using the filter named ``Thin'' located in the first filter wheel. It consists of a substrate of PK-50. This material is a thermal-IR flux blocker opaque for wavelengths $>2.8 \mu$m. At wavelengths $<1.9~\mu$m the background is negligible and the light coming from the target is absorbed by the germanium lens of the air spaced doublet located in the FEU (labeled B in Fig.~\ref{fig:NIRSPEC}). The few observations made in L band have been made using the broad 'KL' filter.

The last component of the background is the internal background of the spectrograph itself. In the case of NIRSPEC, a recently discovered light leak currently adds to the instrument thermal background~\cite{Lopez2020_SPIE}. 

\begin{figure}[ht]
    \begin{center}
    \begin{tabular}{c}
    \includegraphics[width=16cm]{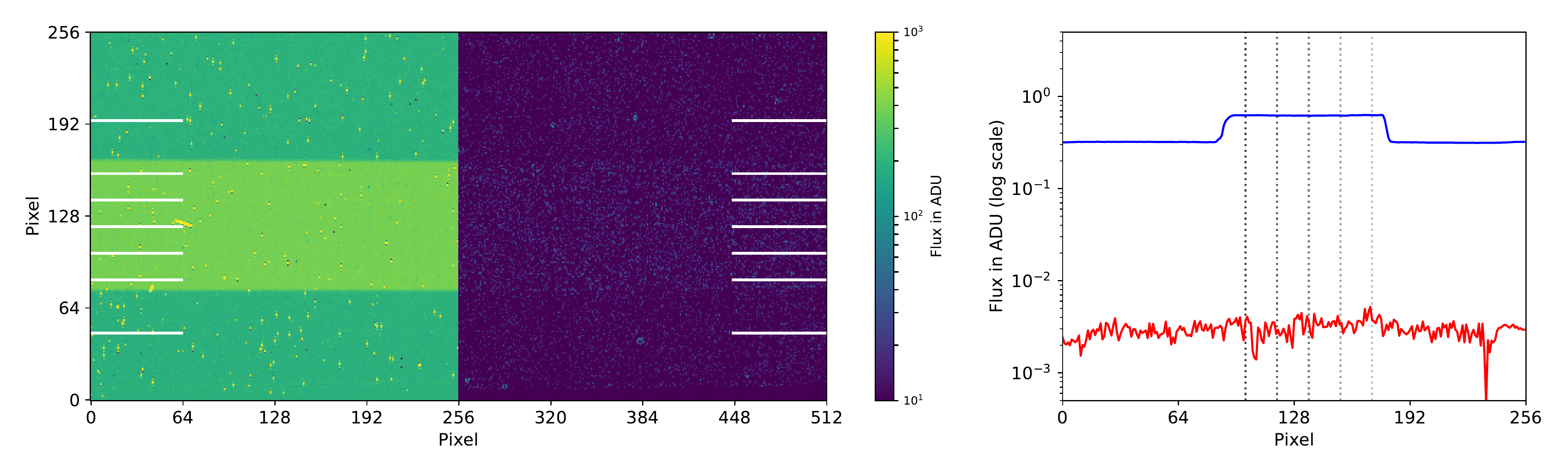}
    \end{tabular}
    \end{center}
    \caption[Background]{\label{fig:background} Left: Central section of order 33 of a Spec frame acquired with no object in the tracking camera FOV (exposure time = 599 sec, Nirspec setup for K band observation, clear sky, elevation = 54.2$^{\circ}$ and no moon). The left side of this image has not been reduced while the right side has been reduced by subtracting a background acquired during the previous night with the same settings and by removing the bad pixels. The horizontal lines of each side indicate the position of the fibers. From top to bottom: C$_A$, S$_1$, S$_2$, S$_3$, S$_4$, S$_5$, C$_B$ (see Fig.~\ref{fig:bundle_diagram}). The slit induced background is visible in the raw data (bright band around the 5 science fibers). Right: Line profiles across order 33. This profile is the median of the profiles associated to all the wavelengths of the order. The profile computed from the raw data is represented in blue while is computed from the reduced data is represented in red.}
    \end{figure}

Figure~\ref{fig:background} shows the central section of one of the orders visible on a Spec frame (left panel) acquired by pointing at a dark region of the sky (no visible target). The left half of the image is not reduced while the right side is reduced by removing a background acquired during the day (no light injected into the fibers) and correcting for hot pixels. The graphic on the right side of the figure presents a vertical line profile of the same order. This profile is the median of the profiles associated to all the wavelengths of the order. The profile computed from the raw data is represented in blue while the profile computed from the reduced data is represented in red. 

%-------------
%-------------

The two first components of the background (sky and fiber emissivity) are not visible in the Fig.~\ref{fig:background}. If they were visible, we would see small peaks at the locations of the fibers in the red trace. Because we do not use exposure times longer than those used to acquire this data during science observations (599 seconds), we can consider these two components as negligible in standard observing conditions. More tests need to be done to determine if the conclusion will be the same in L band, with the telescope pointed at lower elevation or during a night with the moon visible. 

The third component, emitted by warm optomechanical components located in the calibration unit of NIRSPEC (see Fig.~\ref{fig:NIRSPEC}), is visible in the raw data (bright band around the 5 science fibers) but can be removed efficiently by reducing the data. This contribution to the background is not dependent of the observing conditions but can probably be affected by variation of temperature in the AO room. In this case for an AO room temperature of 6.1$^{\circ}$C, we measured an average of 0.5 counts per second for the reddest order (order 39 -- $\lambda \in [2440:2484]$ nm) and less than 0.05 counts per second for the bluest order (order 31 -- $\lambda \in [1943:1977]$ nm).

Regarding the last background component, the thermal background of the instrument, it is combined with all the noise and offsets of the science detectors. This component affects the entire image. Visible as an offset below and above the bright green band on the left side of the image presented by the Fig.~\ref{fig:background}, the sum of these effects contributes 180 counts on the reddest side of the detector and 90 counts on the bluest side. As the temperature inside the cryostat is controlled, we can consider this to be constant. As shown by the Fig.~\ref{fig:background}, this component can be removed efficiently. The optical leak discussed previously is not visible is the sub-image presented in this paragraph. 

\subsubsection{Throughput \label{sec:Throughput}}
\begin{figure}[ht]
    \begin{center}
    \begin{tabular}{c}
    \includegraphics[width=16cm]{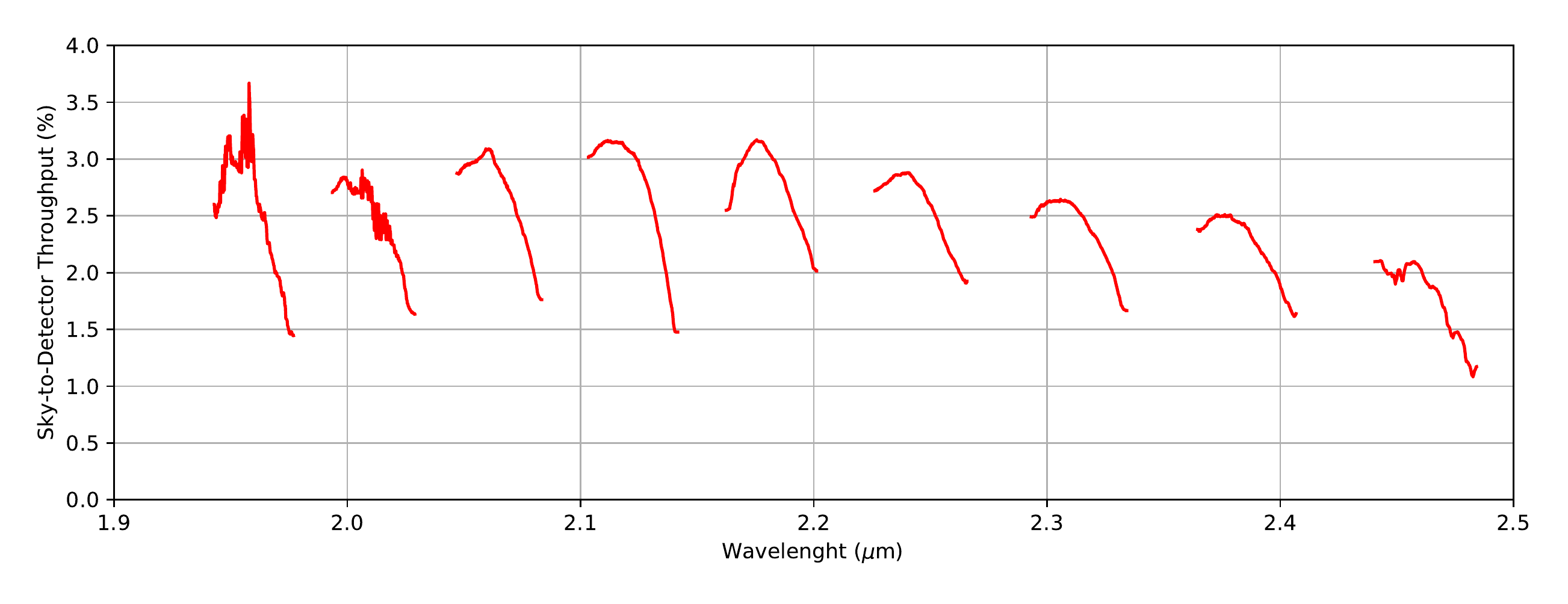}
    \end{tabular}
    \end{center}
    \caption[Scan]{\label{fig:Kappa_And_Throughput} Throughput as a function of the wavelength from sky-to-detector measured on July the 3$^{rd}$ 2020 on the star Kappa Andromedae. For this measurement, the light was injected into science fiber $ScF_{2}$ of the bundle.}
    \end{figure}

Figure~\ref{fig:Kappa_And_Throughput} presents the throughput of the overall system (sky to detector included) as a function of the wavelength measured on July the 3$^{rd}$ 2020 on the star Kappa Andromedae. The throughput is the ratio between the number of photon expected and the number of photons detect by the science detector. Nine orders are visible on this plot. The left one is the 31$^{st}$ order of NIRSPEC while the right one is the 39$^{th}$ order. For this measurement, the light was injected into the second science fiber (labeled $S_{2}$ in Fig.~\ref{fig:bundle_diagram}). This set of data has been acquired during a clear night. according to the Canada France Hawaii Telescope, the MASS seeing was $\leq 0.2$ arc second and a DIMM seeing of  $\approx 0.6$ arc second (both measured for a wavelength of 0.5 $\mu$m). During this measurement the target was at a 64.5$^\circ$ elevation.  

%-------------
%-------------

It can be seen that the peak throughput in K band reaches 3\% (well above our 2\% requirement). The data in the figure represents one of the best results achieved to date. Multiple parameters can affect the throughput: seeing, weather, AO performance, calibration quality. Despite this, over the last few observing runs, we have consistently measured a throughput between 1.5 and 2.5 \% over the K band. 

Neither the Keck II AO system nor the KPIC instrument contains an atmospheric dispersion compensator (ADC). As such, the light of the target is slightly dispersed in the focal plane of the input connector of the bundle for non-zero Zenith angles. For this reason the injection is not uniform across the K band. In the case presented in Fig.~\ref{fig:Kappa_And_Throughput}, the injection was optimized for a selected wavelength (center of the K band). This effect is a function of the elevation of the telescope. To optimize the overall throughput of the system for all the wavelengths, an ADC will be deployed in the phase II of the KPIC project.

We also measured the throughput in L band using Kappa Andromeda at the end of the same night. We measured a preliminary throughput of up to $6\%$ between 3.7 and 3.8 $\mu$m. 

The peak throughput measured on-sky of $3.2\%$ in K band is consistent with the expected value of $3.4\%$, computed assuming 200 nm RMS of residual aberration and taking into account the performance of each of the optics. We have identified several ways to improve the overall performance of the system, which should result in an increase in the quality of the science data in the near future. The following table presents the throughput budget computed for the FIU in both K and L bands.

%-------------
    \begin{table}[ht]
    \begin{center}
    \begin{tabular}{ |c|c|c|c|  }
    \hline
    \multicolumn{4}{|c|}{\textbf{Throughput budget}} \\
    \hline
    \hline
    \textbf{Components} & \textbf{K-band} & \textbf{L-band} & \textbf{Description} \\
    \hline\hline
    Sky & 98\% & 85\% & ---\\
    \hline
    Telescope & 90\% & 90\% & 3 reflective optics\\
    \hline
    Keck II AO optics & 63\% & 63\% & 7 reflective \& 1 transmissive optics\\
    \hline
    Fiber Injection Unit & 77\% & 77\% & 12 reflective \& 2 transmissive optics\\
    \hline
    Strehl & 72\% & 90\% & Assuming 200 nm RMS wavefront error\\
    \hline
    Dispersion \& Pointing loss & 95\% & 95\% & At 30 degrees zenith angle\\
    \hline
    Fiber injection efficiency & 60\% & 60\% & Keck pupil geometry and NA mismatch\\
    \hline
    Bundle & 95\% & 97\% & Material absorption \& Fresnel reflections\\
    \hline
    FEU + Calibration unit & 89\% & 89\% & 2 reflective \& 5 transmissive optics\\
    \hline
    NIRSPEC throughput & 30\% & 30\% & Diffraction efficiency \& optic throughput\\
    \hline
    Filter and background stop & 80\% & 80\% & --- \\
    \hline
    H2RG Quantum efficiency & 95\% & 95\% & --- \\
    \hline\hline
    Total & 3.4\% & 3.7\% & From top of atmosphere \\
    \hline
    \end{tabular}
    \end{center}
    \caption[Throughput budget]{\label{tab:Throughput_budget} Throughput budget computed for the FIU in both K and L bands.}
    \end{table}
%-------------

\section{Calibrations \label{sec:Calibration}}
Injecting light into a SMF, which subtends a very small solid angle on-sky, requires precise calibration of the instrument. In this section, we describe the various calibrations performed before and during the observations.

\subsection{Fiber injection unit calibrations  \label{sec:Cals_FIU}}

\subsubsection{Science fibers position \label{sec:Cals_ScF_Position}}
The science fibers are not visible on any sensors, so their positions are unknown at first. The tracking camera is integral to finding the position of the science fibers and optimizing the coupling of a target. By virtue of the dichroic beamsplitter before the tracking camera (M in Fig~\ref{fig:KPIC_Diagram}), J and H band light can be used to determine the location of a target. To determine the position of the science fibers on the tracking camera, we follow a two step calibration process. The first one is quick but not highly accurate while the second one is time consuming but very accurate. 

\begin{figure}[ht]
    \begin{center}
    \begin{tabular}{c}
    \includegraphics[trim={0 0 0 28}, width=12cm,clip]{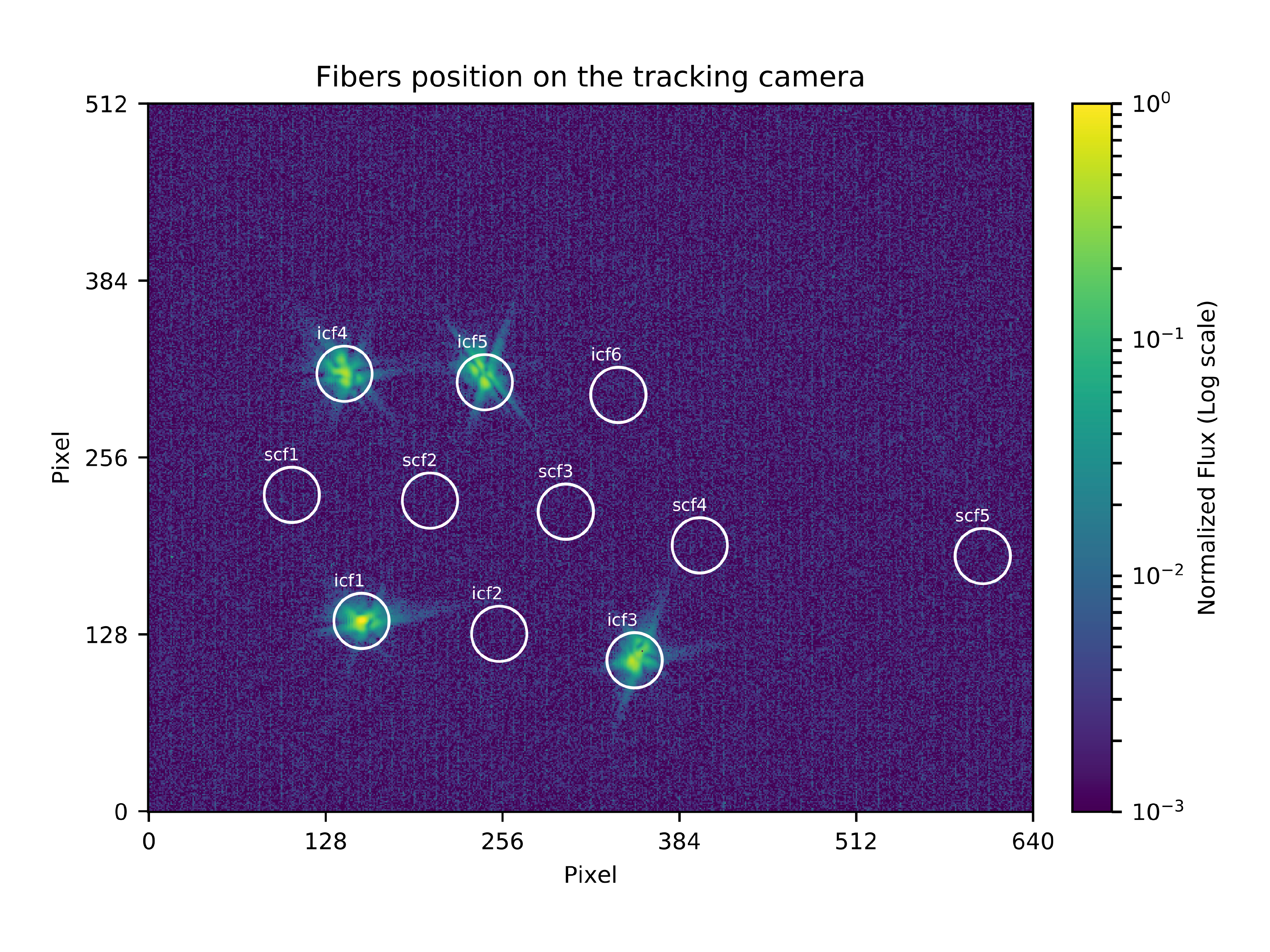}
    \end{tabular}
    \end{center}
    \caption[Scan]{\label{fig:Fibers_Position} Output figure automatically generated by the algorithm used to roughly determine the position of each fiber of the bundle. The "X" shape of the PSFs is due to the corner cube}
    \end{figure}

The first calibration procedure relies on light to be retro-fed through the peripheral calibration fibers in the bundle. An infrared laser is used and injected into four of the six fibers and emitted from the input connector of the bundle (see Sect.~\ref{sec:Bundle}). The light coming from those fibers is collimated by the focusing OAP of the injection module (H in Fig.~\ref{fig:KPIC_Diagram}). A fraction of this light is reflected by the dichroic (M) in the direction of the corner cube (T), which reflects the light in the direction of the dichroic. The light transmitted forms an image on the tracking camera composed of four PSFs (one per retro-fed calibration fiber). By retro-feeding an asymmetric pattern of calibration fibers and by knowing the geometry of the bundle, we can determine the position of the other fibers (science fibers) in the input connector of the bundle. The image presented in Fig.~\ref{fig:Fibers_Position} is the output automatically generated by the algorithm used to perform this calibration procedure. It takes only a few seconds to get a result but the estimated positions are not highly accurate (usually a couple of pixels) for multiple reasons: the shape of the PSFs for the four calibration fibers are altered by the OAP and the corner cube, which makes the position measurement of these fibers inaccurate, even if the geometry of the bundle is precisely known, we do not take into account field distortions at the detector, and the corner cube is not perfect so the incident and reflected beams are not parallel, to name a few. 

%-------------
%-------------

\begin{figure}[ht]
    \begin{center}
    \begin{tabular}{c}
    \includegraphics[width=16cm]{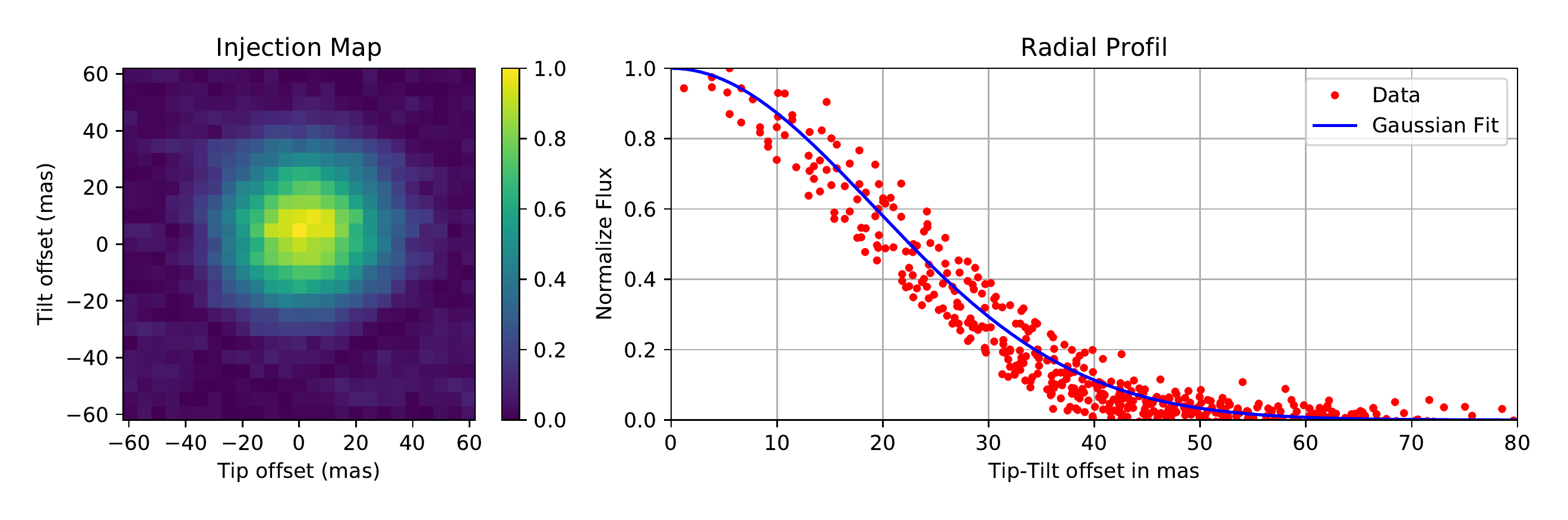}
    \end{tabular}
    \end{center}
    \caption[Scan]{\label{fig:Grid_Scan} Left: injection map obtained by performing a 25 by 25 grid scan. The pitch of the grid is 5 mas in both direction. Right: Radial profile computed from the injection map.}% In both cases, 1 pixel = 8.06 mas.}
    \end{figure}

In order to maximize the throughput of the instrument, we use a second calibration procedure based on a direct measurement of the light injected into the fibers. For this step, we turn on the calibration source of the Keck II AO bench to create a PSF on the tracking camera and then use the TTM of the fiber injection unit (labeled L on Fig.~\ref{fig:KPIC_Diagram}) to align this PSF with one of the science fibers. Then we scan the TTM across the expected location of the fiber and record the flux on the slit viewing camera of NIRSPEC, Scam (see Fig.~\ref{fig:NIRSPEC}) as well as the tracking camera. The SCAM images are used to measure the flux injected into the science fiber while the tracking camera images are used to measure the flux variation of the light source and track the PSF position associated with each flux measurement. In this way, we can build and injection map through the fiber and by fitting the peak, we can determine the TTM position of optimum coupling very accurately and the corresponding PSF position that yields this optimum alignment on the tracking camera at the same time. The left image presented in Fig.~\ref{fig:Grid_Scan}, is the injection map obtained by performing such a scan. The position of each pixel of this injection map correspond to a position of the PSF on the tracking camera and the value associated with each pixel is the flux measured by using Scam normalized by the flux of the calibration source measured by the tracking camera. This calibration method is time consuming. We are currently limited by the minimum exposure time of the Scam detector ($\approx 600$ ms) and its read-out ($\approx 1200$ ms). We tried multiple scan patterns to reduce the time needed to perform this calibration but for the moment, the grid scans are the most accurate and reliable. As shown by the injection map radial profile presented on the right side of Fig.~\ref{fig:Grid_Scan}, knowledge of the position of each fiber is critical if we want to maximize throughput. An offset of one pixel on the tracking camera (8.06 mas) corresponds to a loss of 10\% of throughput, two pixels (16.12 mas) corresponds to a loss of 30\% and three pixels (24.18 mas) to a loss of 55\%. An offset of 50 mas which correspond to the angular resolution of the Keck telescopes at 2$\mu$m correspond to a loss of 97\% of the throughput. This is why we take the time to perform these calibrations properly and why we perform these calibration before each science night even if the bundle has not be disconnected between two observations.

%-------------
%-------------

During the calibrations, the PSFs in H, K and L bands are superimposed. As neither the Keck II AO system nor the first phase of the KPIC project contains an atmospheric dispersion corrector, when on-sky the PSFs of the target in H, K and L bands are superimposed only when observing at Zenith. Therefore, when observing away from Zenith, the location of the PSF on the tracking camera, which operates in H band, will not correspond to the location of the K and L band PSFs in the focal plane of the injection OAP. As such, an additional step is taken during the acquisition sequence, whereby the PSF is offset with the TTM along the elevation axis by the calculated offset expected from the differential atmospheric refraction between the H and science bands (K or L).  This offset is a function of the elevation of the target and the refractive index of the air versus wavelength. To compute the latter, we use an analytical approximation of the refractive index which depends on the temperature, pressure and relative humidity of the air above the primary mirror~\cite{Mathar2007_Journal_of_Optics}. The tracking loop used to align the target with the fiber computes this value and applies the offset at each iteration of the loop (see Sect.~\ref{sec:Acquisition}).

\subsubsection{Tracking camera \label{sec:TC}}
The calibration of the tracking camera is critical because we use it as a reference to inject the light of the target of interest into one of the science fibers. Key to calibrating the camera is understanding the plate scale, the orientation and the field distortion map. We use two different methods to calibrate the tracking camera. The first one can be performed during the day by using NIRC2 as a reference. The plate scale, the orientation and the distortion map of this detector is regularly calibrated and can be considered as reliable. By using a dichroic like the KPIC-pick off (labeled A on Fig.~\ref{fig:KPIC_Diagram}) we can image the PSF of the calibration unit of the Keck II AO bench on both NIRC2 and the tracking camera at the same time. We translate the calibration source in both the X and Y axis and acquire images for each position with both detectors. We compute the position of the PSF in each image. A set of two positions can be used to determine the plate scale of the tracking camera and a set of three positions forming an asymmetric pattern can be use to determine the orientation of the tracking camera with respect to NIRC2. The more sets of data available the more accurate the plate scale and the orientation. The variation as a function of x/y on the detector of plate scale and orientation between those sets of data is used to compute the distortion map of the tracking camera. To model the distortion in X and Y on the detector, we use the methodology of Service et al. 2016~\cite{Service2016_IOP} and use a 5th order bivariate Legendre polynomial. A second calibration method performed during the night is routinely used to refine this. This method consists of observing an object with multiple components (a binary star for example). Because the astrometry of such an object (position angle and separation) is stable during the calibration, we can use images of this object acquired on different parts of the detector to calibrate the distortion map of the tracking camera. To move the object around the detector we steer the TTM in the FIU. Indeed, in the presence of local field distortions, the position angle and separation vary across the detector. The average separation of this target can be used to determine the plate scale and the orientation of the detector. In order to mitigate orbital motion uncertainties of the system observed to perform this calibration, we image the system with NIRC2 and the Cred2 in parallel. We also tried to observe known fields of stars, like M92. However, the FOV of the tracking camera (4.5 x 5.5 arc-second) is too narrow to image many bright stars at the same time (H-mag $\leq$ 10). 

The offsets computed from the calibrations outlined above, are applied to the tracking loop (see Sect.~\ref{sec:Acquisition}) to align the science target with the selected science fiber. We verify at the beginning of each run that we can align the PSF with a sub-pixel resolution to confirm the validity of the tracking camera calibrations. 

\subsubsection{Coupling Optimization}\label{sec:coupopt}
To optimize the coupling, we must reduce the wavefront error in the system in order to maximize the image quality in the focal plane of the input connector of the fiber bundle. However, this plane is not easily accessible and cannot be imaged with a detector so we cannot use any standard focal plane wavefront sensing strategy in this case. 
The solution found to overcome this problem is to use two calibration procedures. The first one is used to optimize the image quality on the tracking camera while the second one is used to calibrate the non-common path aberrations (NCPA) between the two paths of the FIU (tracking camera and injection module). 

To optimize the image quality on the tracking camera we use a phase diversity algorithm called ``image sharpening". This is used in the calibration procedure for some of the Keck facility instruments. This procedure can be run before each science night as it takes only a few minutes to complete. Indeed, even if the version used is not fully automated yet, we never start from a very deformed PSF. We usually start with less than 100~nm RMS of wavefront aberration in the system and consistently reach wavefront errors $<20$~nm RMS by using this procedure.

Once the wavefront error is reduced to maximize the image quality on the tracking camera, we use a Zernike optimization strategy which measures the flux transmitted through the fiber bundle to indirectly reduce the NCPA between the two paths of the FIU. We first select a Zernike mode (we usually start with  the defocus). Then we scan this mode in amplitude by using the Keck AO DM. For each amplitude we perform a tip-tilt scan in order to build an injection map (see Sect.~\ref{sec:Cals_ScF_Position}). This is done incase the PSF move laterally while being defocus (i.e. in case there is cross-coupling between modes). By fitting a 2D Gaussian to each injection map we obtain the optimal injection for each Zernike amplitue. The optimal amplitude for the selected Zernike mode is determined by fitting a second order polynomial on the optimal injection values measured for each amplitude of the selected Zernike mode. We then apply this value to the DM and continue the calibration procedure by scanning the next Zernike mode. This calibration procedure is completed once all the Zernike modes selected have been scanned. Because we currently only scan one mode at a time, we assume the modes in the system are orthogonal, which is a reasonable assumption. In addition, most of the NCPA is in the lowest order modes so we typically dont need to scan beyond 15 Zernike modes. 

\begin{figure}[ht]
    \begin{center}
    \begin{tabular}{c}
    \includegraphics[width=16cm]{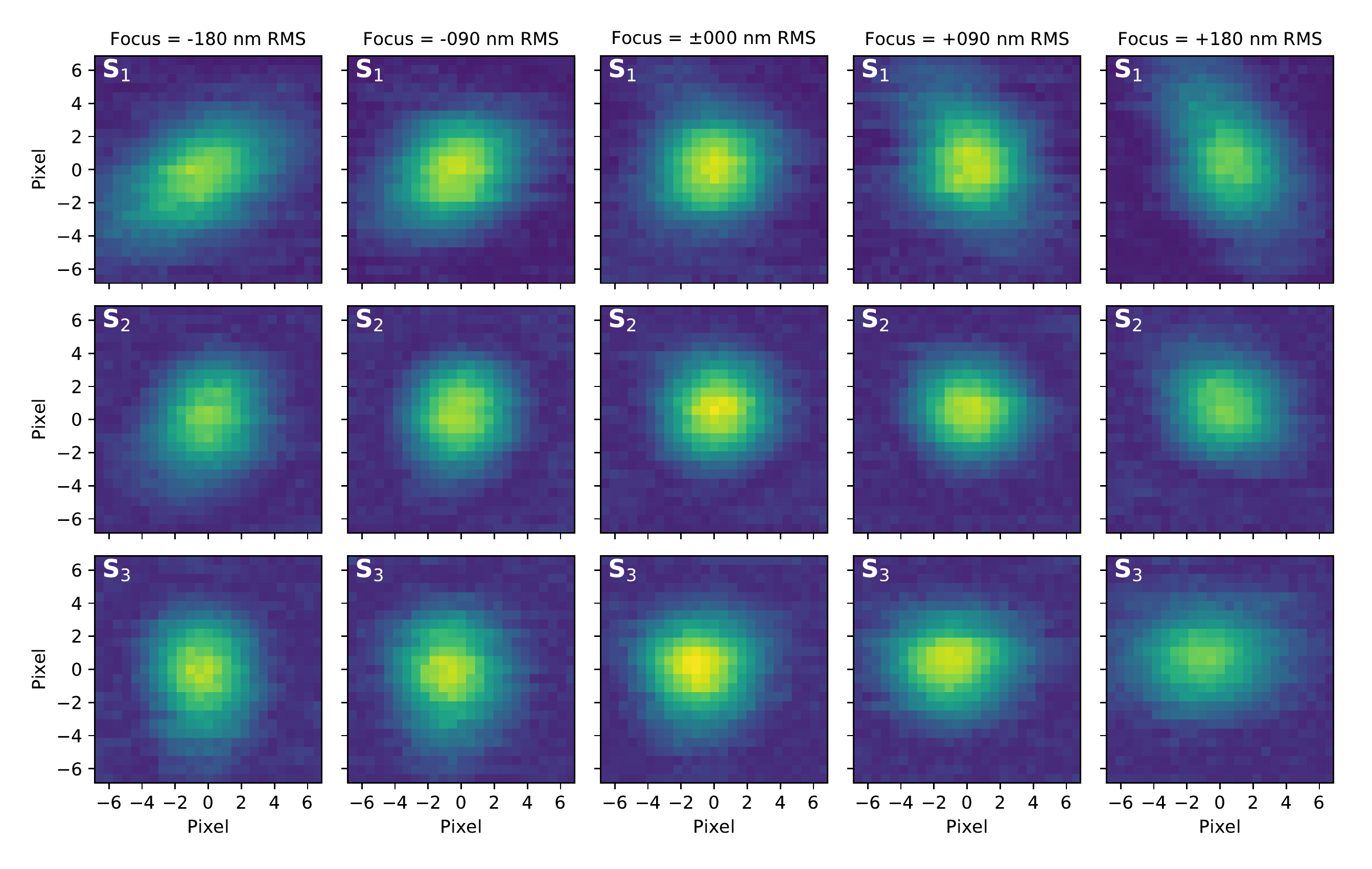}
    \end{tabular}
    \end{center}
    \caption[FocusScanP]{\label{fig:FocusScanP} Measured injection maps from data collected on three fibers: S$_{1}$ (top row), S$_{2}$ (central row), S$_{3}$ (bottom row) and five defocus amplitudes centered on the initial defocus. From left to right: -180, 90, 0, +90 and +180 nm RMS. The flux of the injection map has been normalized by the maximum of all the injection maps measured for those three fibers (63 total).}
    \end{figure}

Figure~\ref{fig:FocusScanP} presents some of the injection maps obtained while optimizing the defocus for the first three science fibers. Each row is associated to a fiber (top row to S$_{1}$, central row to S$_{2}$, and bottom row to S$_{3}$) and each column to a defocus amplitude (from left to right: -180, -90, 0, +90 and +180 nm RMS). In this figure, we can see that the NCPA are not the same for each fiber of the bundle. Indeed, the elongation in the injection maps hints at the presence of astigmatism in the PSF at the location of some of the fibers.  Because this calibration is time consuming and because we mainly use the second science fiber of the bundle on sky, we only run the calibration for this fiber routinely.  

%-------------
%-------------

By using this calibration procedure, we have been able to improve the injection by $\sim13\%$. This result has been obtained by scanning eight Zernike modes (Noll index $\in[4:11]$). The results presented in this paragraph have been obtained by connecting the bundle to a photo-detector using a simple optomechanical assembly. The use of the photo-detector allows us to run the calibration faster but prevents us from running the calibration when the bundle is connected to NIRSPEC. In the future we aim to be able to run this calibration, by using Scam and/or Spec. In addition, we need to improve the algorithm to converge faster. We also want to characterize the NCPA and determined how it has evolved over time in order to determine how often this calibration has to be performed. 

\subsection{Fiber extraction unit calibrations \label{sec:Cals_FEU}}
NIRSPEC can be used on both Nasmyth platforms of the Keck II telescope. As such, the bundle has to be reconnected to the FEU each time the instrument is transported. Even if the re-positioning of the bundle (focus and orientation) is repeatable and the FEU stable over time, we perform basic calibrations of this module before each science night in order to optimize the performance of the instrument. 

The first calibration is used to co-align the fibers contained in the output connector of the bundle with the slit selected for the observation. To move the fibers with respect to the slit, we use the FEU-TTM (labeled C on the Fig.~\ref{fig:NIRSPEC}). To rotate the fibers with respect to the slit we use the rotator mirror contained inside the NIRSPEC cryostat. Usually the rotator does not need to be rotated by more than a degree, because the bundle were clocked to approximately the right orientation by hand initially. 

The second calibration is used to co-align the beam coming from the fiber bundle with the stop used to reduce the background (see Sect.~\ref{sec:NIRSPEC}). This stop is in the pupil plane of the fore-optics. To move the beam with respect to the stop, we translate the bundle in the planes perpendicular to the optical axis using two of the motorized translation stages contained in the FEU (see Sect.~\ref{sec:FEU}). 

The third calibration is used to focus the bundle and adjust the size of the PSFs coming from each fiber of the bundle on the slit. Because NIRSPEC has not been designed to be diffraction-limited, the PSF is smaller than the slit when the bundle is perfectly in  focus. In this case, the spectrum acquired by the science detector of NIRSPEC is undersampled. To overcome this problem, we have the capability to defocus the bundle in order to optimize the sampling of the science data. We adjust the focus of the bundle using one of the motorized translation stages contained in the FEU (see Sect.~\ref{sec:FEU}). At the shortest observed wavelength in a given band, we adjust the focus of the bundle to get 2.4 pixels/FWHM and this produces an R$\sim$35k.

These three calibration procedures are relatively rapid to perform due to the stability of the FEU (usually less than one hour). The adjustments between two consecutive nights are typically minor.

\subsection{NIRSPEC calibrations \label{sec:Cals_NIRSPEC}}
KPIC uses NIRSPEC in non-standard ways, so additional calibration steps on top of the standard ones, are necessary. The days before the observation we take the standard backgrounds and darks with multiple instrument set ups we are likely to use during the science nights. We have the possibility to take extra calibration data after the night if needed but calibration data acquired before the science night can be used to quickly reduce the data during the night to verify if everything is working as expected.

During the days before the observation, a wavelength solution calibration of the science detector is performed by using NIRSPEC's calibration sources (see Fig.~\ref{fig:NIRSPEC}). The different arc lamp sources available in this unit have known emission spectra, that enable wavelength calibration of the instrument. The calibration unit also contains an extended white light source. We use this source to acquire flats for every configuration we plan to use during the night.

The next step of the NIRSPEC calibrations consist on determining where the traces associated with each fiber in the bundle is located on the science detector. The position of these spectra must be known to extract the spectral information of each target observed. In the case of a bright object, the location of the spectra is obvious but this is not the case for faint objects, and so determining where to look for the traces with a calibration source off-sky can be beneficial. To perform this calibration, we set up NIRSPEC, align the fibers of the output connector with the slit using the position determined previously (see Sect.~\ref{sec:Cals_FEU}) and inject the light of the calibration source of the Keck II AO bench in each science fiber, one at a time. For each science fiber, we acquire an image with the science detector. Because the calibration source of the Keck II AO bench is bright and its emission spectra continuous, we can extract the position of the traces associated with each science fiber, across all the orders visible on the detector. Because the position of the spectrum on the science detector and the wavelength solution are dependent on the position of all the opto-mechanical elements located downstream of the bundle, we need to perform this calibration for each instrument configuration we plan to use during the science nights. 

When we observe a target with at least one bright component (K or L magnitude $<$ 10), we repeat this calibration on-sky. By acquiring such a data set on-sky, we can insure that our calibrations were not affected by long term thermal variations as the calibration is taken close to when the science data is acquired. In addition, we can observe the target of interest and perform the calibration by keeping all the opto-mechanical elements located downstream of the bundle fixed, which yields a superior wavelength solution. Finally, the data collected on the main star of the system can be used in the reduction of the science data.

\section{Science data acquisition \label{sec:Acquisition}}

The FIU can be used to observe different kinds of targets. In this section we only described the data acquisition procedure followed to acquire high-resolution spectra of exoplanets and brown dwarfs. In most cases, we cannot image the exoplanet using the tracking camera during an observation. So, in order to blindly align the PSF of an exoplanet with one of the science fibers of the bundle, we need to have an accurate knowledge of the astrometry of the system being observed (position angle and separation). For companions with well-characterized orbits such as the HR 8799 planets for example~\cite{Wang2018_AJ}, we can use the predicted relative astrometry or use data collected during the most recent observation to measure the position. To predict the position of a planet we use the algorithm used by the website "\url{http://whereistheplanet.com/}". If the information is not available or not reliable, we can observe the target with NIRC2 at the beginning of our observing run to measure the astrometric parameters of the companion. 

During the observation, the first step of the acquisition sequence consists of aligning the star with the selected science fiber. To achieve this, we use a tracking loop. Its goal is to align the science target or its host star with a science fiber by controlling the TTM in the FIU (labeled L , on Fig.~\ref{fig:KPIC_Diagram}). It uses as inputs the image from the tracking camera as well as multiple parameters provided by the telescope, the calibration procedures as well as the user. 

\begin{figure}[ht]
    \begin{center}
    \begin{tabular}{c}
    \includegraphics[width=16cm]{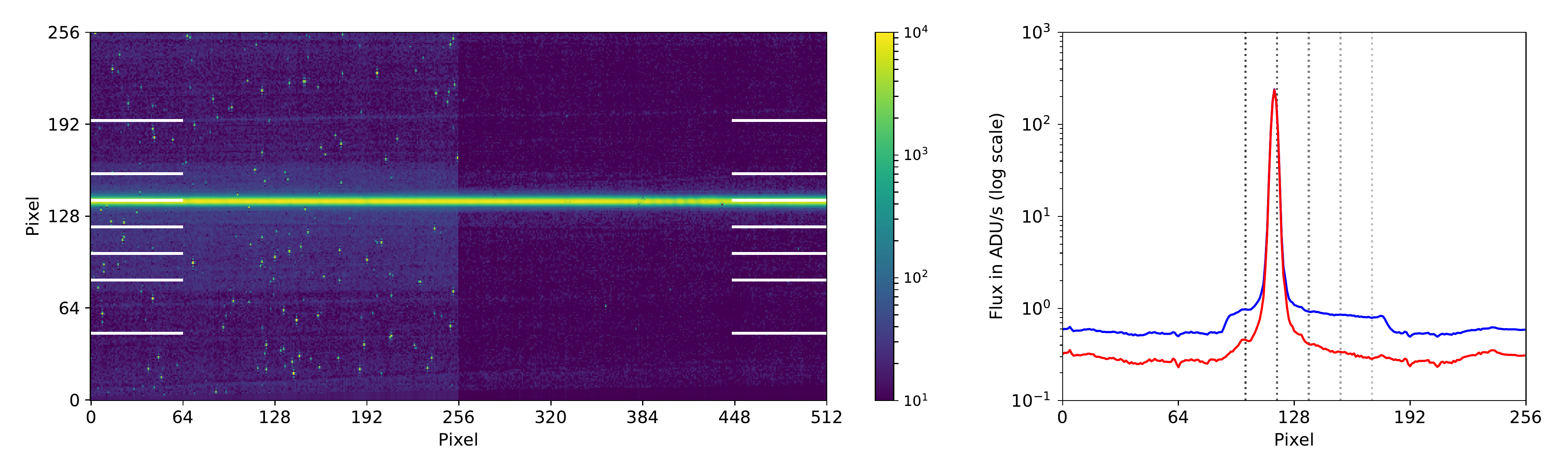}
    \end{tabular}
    \end{center}
    \caption[HR7672A]{\label{fig:HR7672_star} Left: Central section of order 33 of a Spec frame acquired with the star of the HR 7672 system align with the second science fiber (exposure time = 29.5 sec). The left side of this image has not been reduced while the right side has been reduced by subtracting the background and removing the bad pixels. The horizontal lines of each side indicate the position of the fibers. From top to bottom: C$_A$, S$_1$, S$_2$, S$_3$, S$_4$, S$_5$, C$_B$ (see Fig.~\ref{fig:bundle_diagram}). Right: Profiles of order 33. This profile is the median of the profiles associated across all the wavelengths of the order. The profile computed from the raw data is represented in blue while the line profile computed from the reduced data is represented in red.}
    \end{figure}

The algorithm uses tracking camera images to determine the position of the host star on the detector. Using the position of the science fibers determined during the calibration (see Sect.~\ref{sec:Cals_FIU}), the algorithm determines the FIU-TTM offset needed and applies the command. As described in Sect.~\ref{sec:Cals_FIU}, we need to take into account the atmospheric dispersion. This effect is automatically compensated by the tracking loop at each iteration. For a given Zenith angle, temperature, pressure and relative humidity, the offset in the differential atmospheric refraction between H, where the camera tracks and K, where science is conducted, is calculated and applied along the elevation axis (axis of refraction). At this point we usually acquire data using the science detector of NIRSPEC on the host. These data are used to measure the throughput of the instrument and to reduce the science data acquired later in this sequence. Figure~\ref{fig:HR7672_star} presents data acquired with the star of the HR7672 system aligned with the second science fiber (labeled S$_{2}$ on Fig.~\ref{fig:bundle_diagram}) on July 3$^{rd}$, 2020. The left panel presents the central section of the order 33 ($\lambda \in$ [2047:2083] nm) before and after reduction. The graph presented on the right side of the figure presents the median line profile of this order. The blue curve is the line profile computed from the raw data while the red one is computed from the reduced data. The spectra of the star, aligned with the second science fiber of the bundle, is clearly visible on both the Spec frame and the graph presented by this figure. 

%-------------
%-------------

\begin{figure}[ht]
    \begin{center}
    \begin{tabular}{c}
    \includegraphics[width=16cm]{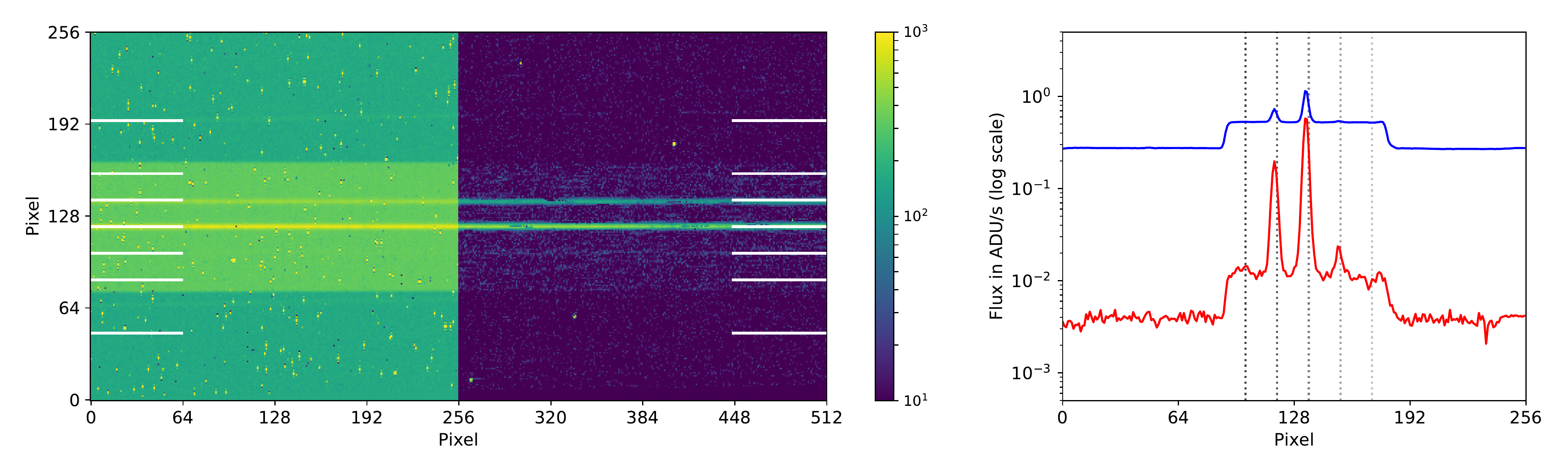}
    \end{tabular}
    \end{center}
    \caption[HR7672B]{\label{fig:HR7672_companion} Left: Central section of the order 33 of a Spec frame acquired with the companion of the HR 7672 system align with the second science fiber (exposure time = 599 sec). The left side of this image has not been reduced while the right side has been reduced by subtracting the background and removing the bad pixel. The horizontal lines of each side indicate the position of the fibers. From top to bottom: C$_A$, S$_1$, S$_2$, S$_3$, S$_4$, S$_5$, C$_B$ (see Fig.~\ref{fig:bundle_diagram}). Right: Profiles of the order 33 computed by averaging the profile associated to each wavelength of the order. The profile computed from the raw data is represented in blue while the profile computed from the reduced data is represented in red.}
    \end{figure}

After aligning the star with the selected science fiber and the spectral data collected, we use the relative astrometry of the companion to offset the star in order to align the science target with the fiber. This offset is applied by providing the astrometric parameters of the target (position angle and separation) to the tracking loop. After correcting the astrometric parameters to take into account the field distortion of the tracking camera, the tracking loop computes the offset needed and sends a command to the FIU-TTM to apply the offset. We verify at the beginning of each run that we can apply such offsets with a sub-pixel accuracy by observing several binary stars with both components visible on the tracking camera. At this point, integration on the science target (the companion) begins. In the case of faint targets (K or L magnitude $\geq$ 13), we acquire multiple 10 min exposure frames with the science detector of NIRSPEC. Figure~\ref{fig:HR7672_companion} presents, on its left side, a section of order 33 extracted from one of the science frames acquired  the HR7672b system aligned with the second science fiber. The graph presented on the right side of the figure presents the median profiles of this order. The blue curve is the profile computed from the raw data while the red one is computed from the reduced data. 
Three peaks are visible in the line profile, which indicate the position of the fiber traces. The one associated with the second science fiber contains a combination of the light coming from the science target, the speckle field of the star and the multiple backgrounds described in Sect.~\ref{sec:Background}. The line associated with the third and fourth science fiber contains the light of the the speckle field of the star and the multiple backgrounds (see Sect.~\ref{sec:Background}). We cannot adjust the intensity of the light coming from the speckle field of the star in the science fiber aligned with the companion as the separation between the companion and its host start is constant. However, it can be adjusted for the other fiber by moving the star around the science target using the rotator of the telescope. The offset applied to the rotator is automatically taken into account by the tracking loop.

%-------------
%-------------

During the observation of a faint target, we regularly acquire data with the star of the system aligned with the selected fiber (every hour or 6 science frames) in order to get enough calibration data and to make sure the injection into the fiber is high and stable by measuring the throughput of the system. 

\section{On sky performance and first results \label{sec:First_Results}}

To demonstrate the science capabilities of the system, we observed HR 7672 B on September 28 2020, a bright ($K = 13$~mag) early L-type brown dwarf that is expected to have a similar spectral signature as many of our more challenging targets~\cite{Liu2002_ApJ}. Residing 0.64 arcsec from the star at the epoch of observation~\cite{Crepp2012_APJ}, the companion also resides at comparable separations from the star as our exoplanet targets. We obtained ten 600~seconds exposures of the brown dwarf placed on science fiber S$_2$, resulting in a total integration time of 100~minutes. The data were reduced using a preliminary version of the KPIC data reduction pipeline. The thermal background of the instrument was subtracted using 8 hours of backgrounds taken during the day with the same instrument setup. The A0 standard star ups Her was observed at the start of the night to calibrate the position and width of all of the fiber traces. We used this trace information to extract spectra from all of the science fibers of the HR 7672 exposures, resulting in on-axis observations of both HR 7672 A and B on S$_2$. The extracted spectra were then averaged in time, removing bad pixels due to cosmic ray contamination. The M-giant HIP 81497 was observed in each fiber to compute the wavelength solution of each fiber. The A0 star zet~Aql was observed immediately before the HR 7672 observations to measure the total end-to-end spectral response, including the telluric absorption profile. 

\begin{figure}[ht]
    \begin{center}
    \begin{tabular}{c}
    \includegraphics[width=1\linewidth]{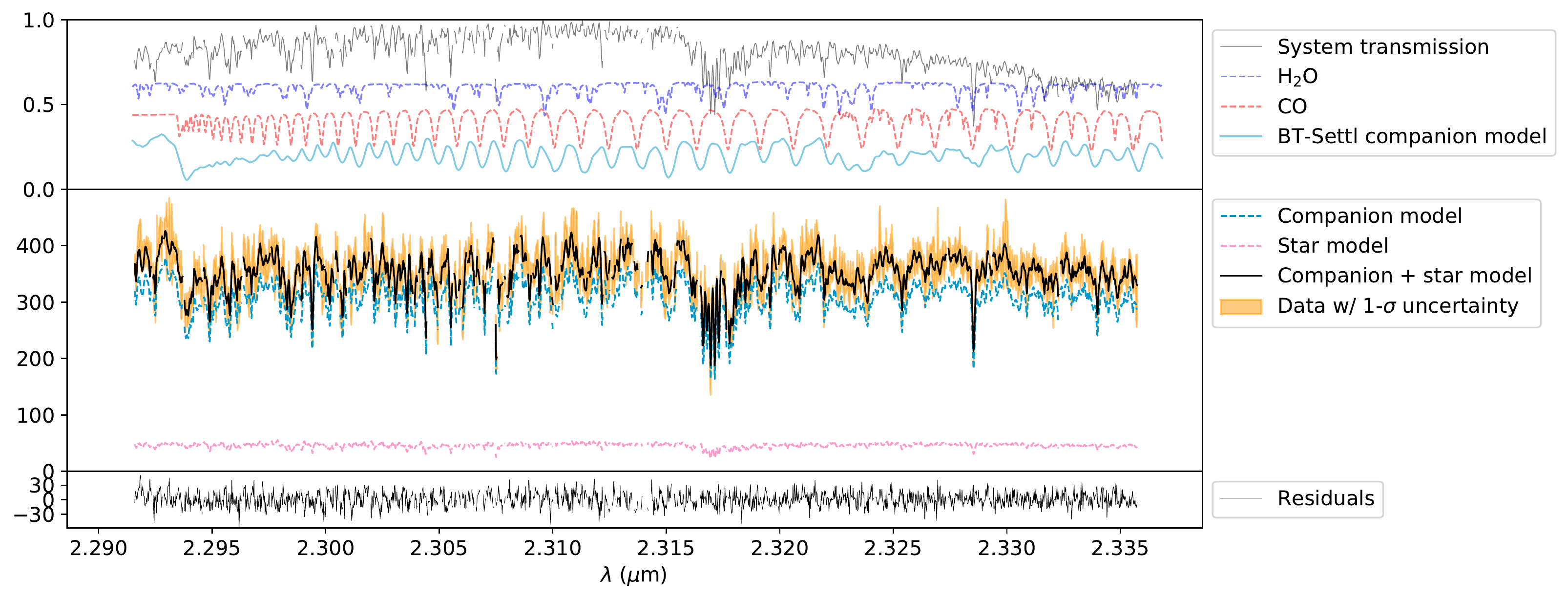}    
    \end{tabular}
    \end{center}
    \caption[spectrum]{\label{fig:HR7672_spectrum} Spectrum of HR 7672 B (37$^{th}$ NIRSPEC order). The middle panel includes the extracted spectrum of the science fiber including its 1-$\sigma$ uncertainties (orange contours). The spectra were high-pass filtered but plotted with an offset corresponding to the continuum median value. The best-fit forward model of the data is shown as a black line. Because the spectrum of the companion is contaminated by the glare of the star, we jointly fit an on-axis starlight spectrum with a model of the planet. Both components of the model, which include telluric lines, are shown separately as dashed lines. The companion is modeled with a BT-SETTL template assuming a $1800\,\mathrm{K}$ temperature and a $\log(g)=5.0$ surface gravity.
    The top panel features spectral templates that can be used to identify spectral lines in the data. The system transmission includes: the transmission of the atmosphere and the instrument.  The BT-Settl model includes spin broadening of the companion, while the molecular template is only broadened to the resolution of the instrument. 
    The lower panel shows the residuals of the fit. }
    \end{figure}

\begin{figure}[ht]
    \begin{center}
    \begin{tabular}{c}
    \includegraphics[width=16cm]{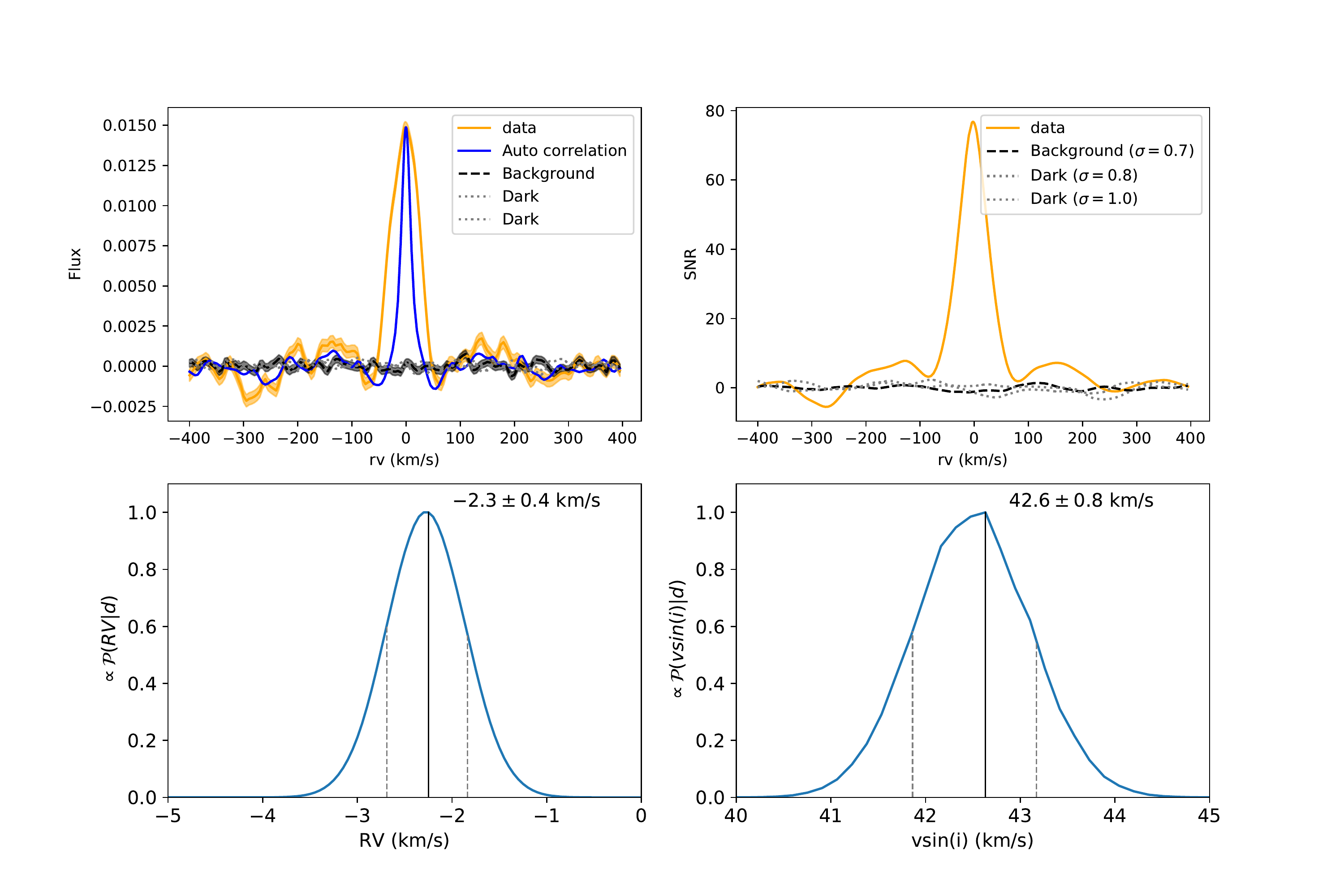}
    \end{tabular}
    \end{center}
    \caption[CCF]{\label{fig:HR7672_CCF} Detection of HR 7672 B using the four reddest NIRSPEC orders (Order 36-39). The top left panel is a type a cross correlation analysis. It is calculated using the forward model framework illustrated in Figure~\ref{fig:HR7672_spectrum} as the estimated flux and associated uncertainty of the brown-dwarf companion as a function of the radial velocity (RV) shift of the model assuming a null spin for the companion. Fictitious fibers that only include background and dark flux are also analyzed with the same model as a way to sample the noise. The top right panel is the signal to noise (S/N) of the detection as a function of RV assuming the best fit spin ($v\sin(i)=42.6\,\mathrm{km/s}$) for the model. The S/N is calculated as the estimated amplitude of the companion divided by its estimated uncertainty featured in the top right panel. The bottom two panels are the marginalized posteriors for the radial velocity and the spin of the companion respectively. The estimated radial velocity of the companion is consistent with the prediction from relative astrometry measurements of the companion and the radial velocity of the host star.}
    \end{figure}

The spectral response, wavelength solution, on-axis observations of S$_2$, and planetary atmosphere models were used to construct forward models of the on-axis spectra of the brown dwarf companion. We used a BT-SETTL atmosphere model with an effective temperature of 1800~K and a $\log(g)=5.0$ for the planet spectrum~\cite{Allard2012}. We varied the radial velocity of the planet, rotational broadening, and overall flux of the planet in fitting the forward model to the data. A more detailed description of the data extraction and fitting technique will be discussed in a subsequent paper. Figure~\ref{fig:HR7672_spectrum} presents the best-fit model and data from order 37 of the spectrum of HR 7672 B. The best-fit forward model matches the data very well, as can be seen in visual inspection. To better visualize the detection, we show the cross correlation function detection of HR 7672 B in the top row of Fig.~\ref{fig:HR7672_CCF}. There is a clear detection of the brown dwarf and is consistent with previous $R\sim1400$ spectra at the same wavelength~\cite{Liu2002_ApJ}. The posterior distributions for the radial velocity of the planet and its rotational broadening are shown in the bottom row of Fig.~\ref{fig:HR7672_CCF}. The vsin(i) measured is (42.6$\pm$0.8)~km/s, and is the first detection of rotational broadening for this companion. The detection and spin measurement of a companion just 0.64~arcsec from its host star demonstrates the HDC capabilities of KPIC.

%-------------
%-------------

%-------------
%-------------

\section{Conclusion \label{sec:Conclusion}}
The fiber injection unit, part of the KPIC demonstrator, is a powerful tool to acquire high spectral resolution data on faint objects surrounding stars, faint isolated objects and close binaries. Deployed at the summit of Maunakea in the fall 2018, the FIU has completed commissioning and has begun regular science operations earlier this year. Numerous known stellar systems have been imaged already and science data acquired. These data, currently being reduced and analyzed, will be published in subsequent papers. After multiple months of science exploitation, the second phase of the project will be deployed in order to improve the overall performance of the instrument. The hardware, software and observing strategies developed, tested and used by both phases of this project will help to prepare future instruments on both ground and space based telescopes.

\acknowledgments 
This work was supported by the Heising-Simons Foundation through grant \#2019-1312. It has been previously submitted as an SPIE proceeding \cite{Delorme2020_SPIE}. W. M. Keck Observatory is operated as a scientific partnership among the California Institute of Technology, the University of California, and the National Aeronautics and Space Administration (NASA). The Observatory was made possible by the generous financial support of the W. M. Keck Foundation. The authors wish to recognize and acknowledge the very significant cultural role and reverence that the summit of Mauna Kea has always had within the indigenous Hawaiian community. We are most fortunate to have the opportunity to conduct observations from this mountain.

%%%%% References %%%%%
% Bibliography data contains in Bibliography.bib
\bibliography{main}   
% Makes bibtex using spiejour.bst
\bibliographystyle{spiejour}

%\end{spacing}
\end{document}